# Watching Single Unmodified Enzymes at Work

Cuifeng Ying,[1, 2] Edona Karakaçi,[1] Esteban Bermúdez-Ureña,[1,5] Alessandro Ianiro,[1] Ceri Foster,[1] Saurabh Awasthi,[1] Anirvan Guha,[1] Louise Bryan,[1] Jonathan List,[1] Sandor Balog,[1] Guillermo P. Acuna[3], Reuven Gordon[4], Michael Mayer[1,*]

[1]Adolphe Merkle Institute, University of Fribourg, Chemin des Verdiers 4, CH-1700 Fribourg, Switzerland.

[2]Advanced Optics and Photonics Lab, Department of Engineering, School of Science and Technology, Nottingham Trent University, United Kingdom

[3]Department of Physics, University of Fribourg, Chemin du Musée 3, CH-1700 Fribourg, Switzerland.

[4]Department of Electrical and Computer Engineering, University of Victoria, Victoria, British Columbia V8P 5C2, Canada

[5]Present address: Centro de Investigación en Ciencia e Ingeniería de Materiales and Escuela de Física, Universidad de Costa Rica, San José, 11501, Costa Rica

Corresponding Author Email: michael.mayer@unifr.ch

**Abstract**
Many proteins undergo conformational changes during their activity.[1,2] A full understanding of the function of these proteins can only be obtained if different conformations and transitions between them can be monitored in aqueous solution, with adequate temporal resolution and, ideally, on a single-molecule level.[3–5] Interrogating conformational dynamics of single proteins remains, however, exquisitely challenging and typically requires site-directed chemical modification combined with rigorous minimization of possible artifacts.[6–8] These obstacles limit the number of single-protein investigations.[4,9] The work presented here introduces an approach that traps single unmodified proteins from solution in a plasmonic hotspot[7] and makes it possible to assign changes in refractive index to changes in protein conformation while monitoring these changes for minutes to hours with a temporal resolution at least as fast as 40 μs. The resulting single molecule data reveals that adenylate kinase employs a hidden enzymatic sub-cycle during catalysis, that citrate synthase populates a previously unknown intermediate conformation, which is more important for its enzymatic activity than its well-known open conformation, that hemoglobin transitions in several steps from its deoxygenated and rigid T state to its oxygenated and flexible R state, and that *apo*-calmodulin thermally unfolds and refolds in steps that correspond to conformational changes of individual protein domains.

**Introduction**

The advent of high-resolution cryo-EM methodology not only accelerates the elucidation of novel protein structures, but also reveals various conformations of the same proteins at an unprecedented pace.[10–12] To study the transitions between these conformations, broadly accessible and rapid approaches for monitoring conformational dynamics in aqueous solution are needed.[6,7] Ideally, these approaches should have the capability to interrogate single molecules to circumvent limitations imposed by ensemble measurements such as averaging and masking of transient intermediates.[3,8,13] Single molecule fluorescence resonance energy transfer (smFRET) and single molecule force spectroscopy (smFS) are the most successful methods for interrogating the conformational dynamics of individual proteins[6–8,14,15] and both techniques reveal fascinating insight into the dynamics of the conformational states of single proteins during enzymatic activity,[16] stepping of motor proteins,[17] unfolding of individual protein domains,[18,19] protein-protein interaction[20] and the activation of receptors, transporters and ion channels.[3,21]

While both methods provide ever more detailed insights into protein dynamics, their application can pose significant challenges.[4,9] For instance, both methods require chemical modification of proteins.[6,7] In the case of smFRET, at least two complementary fluorophores must be attached to appropriate locations on the protein such that the expected conformational change leads to a change in distance between the FRET pair.[3,6] In the case of smFS, two locations of a single protein must be attached to a macroscopic surface, microbead or cantilever.[7,8] These tasks typically require prior knowledge about the amino acid sequence and structure of the protein, site-directed modification of the protein, and rigorous minimization of artifacts such as non-specific adsorption, and the attachment of multiple tethers, fluorophores or proteins.[6–8] Additional challenges of smFRET include photobleaching,[3] minimization of background fluorescence, distinguishing fluorophore dynamics from conformational dynamics,[13] and quenching of fluorophores[22] by metalloproteins.[23] The strength of smFRET studies, however, lies in their ability to quantify the change in distance with high sensitivity and accuracy and to afford a temporal resolution that can reach tens of microseconds.[24,25] For smFS, the challenges include the low yield of attachment of a single protein to two macroscopic objects,[7] tip contamination,[7] possible artifacts from the application of a unidirectional force on the dynamics of an active protein,[8,15] noise and artifacts induced by molecular tethers,[7] as well as limited temporal resolution that is typically slower than 100 μs.[7] The strengths of smFS are its ability to manipulate a single protein in a controlled manner with exquisite precision, to provide a direct measurement of changes in distance and force, and to examine a single protein for long times.[7]

With these strengths and weaknesses of the most successful techniques in mind, a single molecule method for studying conformational dynamics of proteins that interrogates native proteins without modification, with a time resolution that can resolve large domain conformational changes, and with the ability to monitor a single protein continuously without photobleaching or tip contamination for extended times would provide complementary and enabling benefits. Interrogating unmodified proteins would accelerate discovery, reduce the complexity of experiments and make them accessible to a broad range of investigators. Most importantly, the ability to interrogate native proteins without modification in physiological solution would rule out possible uncertainties about the effect of unidirectional forces, fluorescent labels or tethering on conformational dynamics during protein activity.[8,13]

Here we show that a plasmonic nanostructure with a double nanohole (DNH) in a thin film of gold makes it possible to trap individual unmodified proteins for hours in a plasmonic hotspot[26,27] and that the transmitted light through this DNH structure makes it possible to monitor conformational dynamics of the trapped protein in physiologic aqueous solution.[28] We demonstrate with four representative proteins that changes in protein conformation change the local refractive index in the hotspot area and lead to concomitant changes in transmitted light making it possible to assign transmission levels to various protein conformations. The approach provides previously inaccessible insights into the conformational dynamics of single proteins during allosteric conformational change, thermal denaturation of individual protein domains, and conformational dynamics of a monomeric and a dimeric enzyme during their catalytic cycle.

## Results and Discussion

Figure 1a shows the experimental setup used to trap and interrogate individual proteins. This setup is essentially an optical tweezer with the capability of trapping nanoscale objects.[26,29] Trapping such small objects requires exceptionally steep gradients in electrical field at a defined location.[28,29] Figure 1b shows that illumination of a double nanohole (DNH) structure in a thin film of gold with a laser of suitable wavelength creates an optical plasmonic hotspot with at least 20-fold increased electric field in the narrowest part of the DNH structure. Since the electric field enhancement in this hotspot decays within a distance of less than 20 nm, this setup provides a sufficiently steep field gradient for trapping nanoparticles and single proteins.[26,30] To do so, the material properties of these particles have to provide a contrast in refractive index relative to the surrounding aqueous solution.[29] This refractive index contrast not only affects the magnitude of the restoring force towards the hotspot and hence the stiffness of the trap but also the intensity of the light transmitted through the DNH structure.[29] To monitor dynamic processes that affect the refractive index or size of objects in the hotspot, an avalanche photodiode (APD) detects changes in transmission with a temporal resolution as fast as 20 ns. Figure 1c predicts by Finite Difference Time Domain (FDTD) simulation that the experimental configuration used here leads to a decrease in transmitted light when the refractive index of the trapped particle is larger than the one of the surrounding fluid (see Supplementary Information SI-1 and SI-2 for details), as observed experimentally by others.[31] In addition, Figure 1d shows that an increase of the diameter or the refractive index of the particle in the hotspot results in a further reduction of transmission. Figure 1e verifies these predictions experimentally: trapping 22 nm polystyrene nanoparticles with a refractive index of ~1.58 reduced the transmission more than 22 nm silica particles with refractive index of ~1.45. Figure 1f provides complementary verification by trapping a single hemoglobin protein in a deoxygenated buffer (we confirmed trapping of a single protein by determining the trap stiffness as a function of protein size, see Supplementary Information SI-3). As expected from the simulations, replacing the aqueous buffer solution in the plasmonic hotspot by trapping a hemoglobin protein decreased transmission due to an increase of the refractive index in the hotspot.[31] Since the refractive index of solutions containing deoxy-hemoglobin increases when the protein is exposed to oxygen,[32] we investigated whether oxygenation of a single, trapped deoxy-hemoglobin molecule *in situ* further reduces transmission. Figure 1f,g shows that, indeed, the transition from a single trapped deoxy-hemoglobin protein in its T-state to oxy-hemoglobin in its R-state resulted in a strong, stepwise decrease in transmission, resembling a transition predicted by molecular dynamics simulations by Karplus et al.[1] The total duration of this transition of 80 ms is composed of the dwell times of each transmission level and the durations of the changes between these levels (Fig. 1g). We hypothesize that the dwell times are determined by the gradually increasing oxygen concentration as the fluidic system replaced the deoxygenated buffer with oxygenated buffer, and by the on-rate of binding of oxygen to hemoglobin, while the durations for switching between different transmission levels (10.1 ms, 9.2 ms, 6.7 ms, 4.2 ms and 2.9 ms in sequence from T to R) are determined by the conformational changes of hemoglobin subunits from the T to the R state. Ensemble measurements by Hofrichter et al. reported a rate constant for the full quaternary conformational change of trout hemoglobin from the T state to the R state of $14 \pm 5$ $s^{-1}$ at 20 °C, corresponding to a transition time of 50 - 110 ms in agreement with the durations observed here.[33] Figure 1f also shows that the oxygenated state of hemoglobin resulted in additional fluctuations in transmission with higher amplitude and lower frequency than the fluctuations of deoxy-hemoglobin. Analysis of the autocorrelation function of the transmission signal in Figure 1f revealed that trapped deoxy-hemoglobin resulted in a time constant of 0.6 ms consistent with the expected value of a protein with a molecular weight of 64.5 kDa (see Supplementary Information Figure S2 and Figure S3). Oxygenation introduced an additional time constant of 7.6 ms as a result of additional fluctuations of the transmitted light at lower frequency (Supplementary Information SI-4). These fluctuations with increased amplitude are consistent with a trapped hemoglobin protein, which, upon oxygenation, changes from a tense (T) conformation that is stabilized by four salt bridges to a relaxed (R) conformation that is more flexible.[1,23,34,35] Control experiments performed by infusing oxygenated buffer solutions in a DNH with an unoccupied hotspot, or with a hotspot that trapped methemoglobin whose $Fe^{3+}$ ion in its heme groups cannot bind oxygen,[36] did not result in changes in the magnitude or frequency of transmission fluctuations (see Figure S4 and S5).

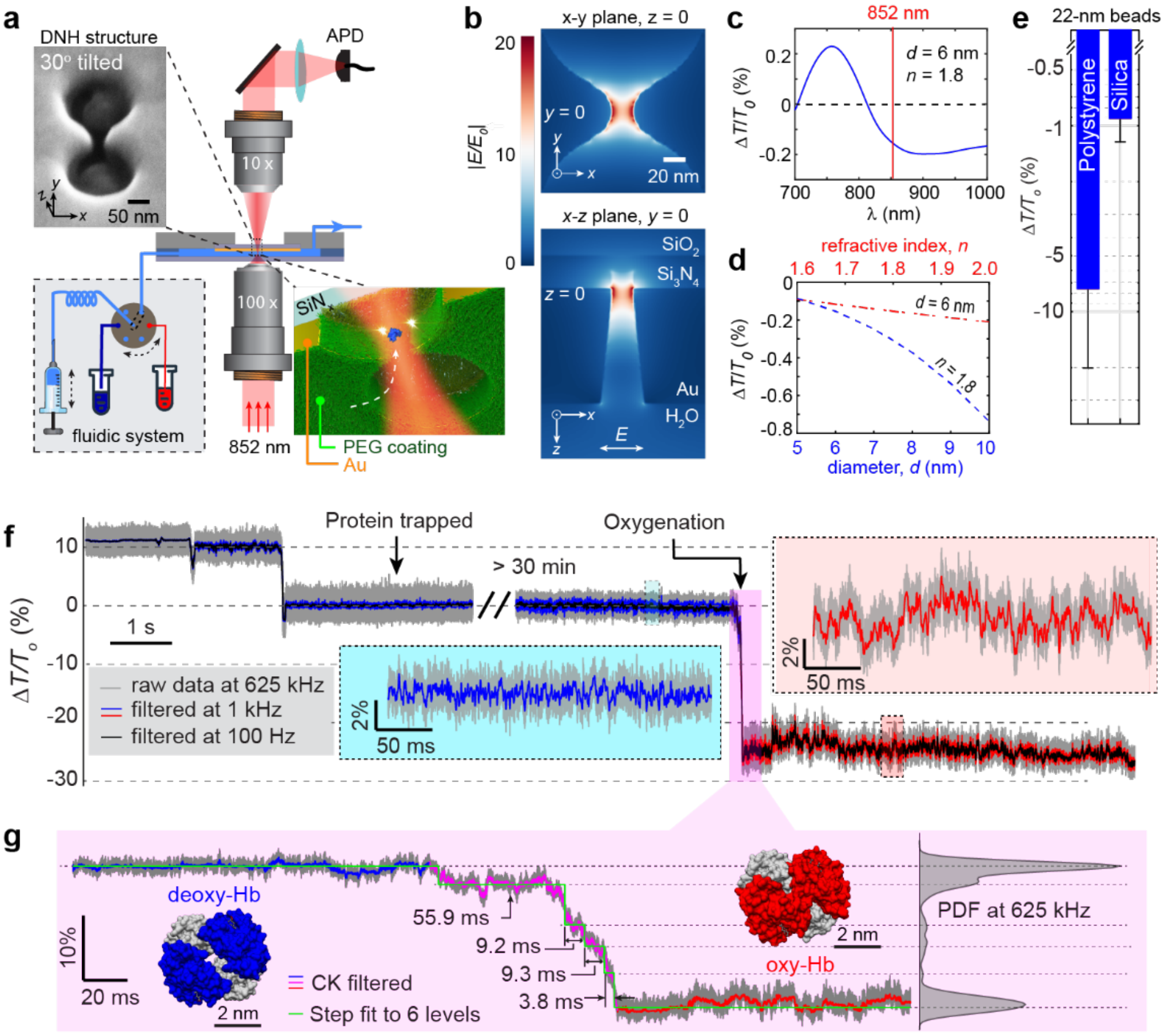

**Figure 1.** Monitoring conformational dynamics of single proteins from modulations of transmitted light through a nanoplasmonic optical tweezer with a trapped protein. **a**, Schematic of the plasmonic optical tweezer platform. A fluidic system injects buffers with various substrates or inhibitors into the flow cell. Upper-left inset: Scanning electron microscopy (SEM) image of a double-nanohole (DNH) structure taken at an angle of 30º above the plane of the DNH. Bottom-right inset: illustration of a trapped protein (blue) in the plasmonic hotspot (white) with a protein resistant coating (green). **b**, Simulated spatial distribution of field enhancement in the DNH at a wavelength of 852 nm obtained by FDTD (see Supplementary SI-1 Materials and Methods and SI-2 for details). The top portion of the panel shows the *x-y* plane at the interface of the gold and Ti layers ($z$ = 0). The bottom portion shows the *x-z* plane at $y$ = 0. The incident laser was linearly polarized along the $x$-axis. **c**, Simulated change in transmission, $\Delta T/T_0$, as a function of wavelength after trapping a particle with a diameter, $d$, of 6 nm and a refractive index, $n$, of 1.8. **d**, Simulated $\Delta T/T_0$ (at $\lambda$ = 852 nm) changes with increasing particle size (blue, $n$ = 1.8) and refractive index (red, $d$ = 6). **e**, Measured change in transmission through a DNH upon trapping single polystyrene ($n$ = 1.576) beads and single silica ($n$ = 1.453) beads, both with a diameter of 22 nm. **f**, Transmission through a DNH as it traps a hemoglobin (Hb) protein, which then transitions from its T state (deoxy-Hb, blue) to its R state (oxy-Hb, red) in response to the introduction of an oxygen-rich buffer. Absorbance spectra confirmed that Hb was in the T state in the deoxygenated buffer, and in the R state in the oxygenated buffer (see Supplementary Information Figure S6). **g**, Stepwise transition while a single deoxy-Hb is exposed to oxygen. The green line is the result of a step-fit to the 6 intensity levels in the probability density function (PDF) of the transmission intensity. Durations for switching between adjacent transmission levels varied between 0.4 and 82 ms from experiment to experiment. Crystal structures depict deoxy-Hb (1HGA, blue) and oxy-Hb (1BBB, red).

The results shown in Figure 1f,g suggest that the nanoplasmonic tweezer setup used here is able to report conformational changes of unlabeled, entirely native proteins on a single molecule level. Since the refractive index of a material is determined by the polarizability of its constituent molecules according to the Lorentz-Lorenz equation,[37] we henceforth refer to polarizability in the context of single molecule experiments. As the polarizability of molecules increases, for instance due to an increase in size (Fig. 1d,e)[38] or an elongation in shape,[28,38] the refractive index of materials composed of these molecules or of solutions of these molecules increases as well.[39] This molecular perspective provides a possible explanation for the reduction in transmission of trapped hemoglobin upon oxygenation. When deoxy-hemoglobin changed conformation from its compact T state to its slightly flattened and elongated R state,[35] the resulting increase in polarizability is consistent with the observed decrease in transmission. The data in Figure 1f provide the first single molecule data on monitoring hemoglobin conformation before, during and after oxygenation; smFRET experiments for monitoring this transition are not possible due to quenching of fluorescence by hemoglobin's heme groups.[23]

To confirm that the experimental setup in Figure 1 can indeed monitor conformational changes of single molecules while excluding possible contributions from binding and unbinding of ligands such as oxygen, we turned to thermal denaturation of single proteins. We chose to investigate unfolding of the calcium-binding protein calmodulin because in the absence of calcium ions, calmodulin (*apo*-CaM) undergoes well-known unfolding transitions of individual domains at various temperatures between 41 and 58 °C,[40–43] while the calcium-bound form of calmodulin (*holo*-CaM) is stable up to at least 78 °C and provides an ideal control experiment.[44]

Figure 2a shows the change in transmission through the DNH upon exposing a single trapped *apo*-CaM protein to stepwise increases in temperature (see Supplementary Information, Figure S7 in SI-8 for the ability to control the temperature in the plasmonic hotspot by the laser intensity). At 41.5 °C, apo-CaM showed sporadic and small fluctuations from a baseline transmission level to a second level with reduced transmission. At 46 °C, this second level was the most probable level. This transmission level remained stable for minutes at 47 °C without significant additional fluctuations (see Figure S8). At 51.5 °C, we started to observe additional transient reductions in transmission; these fluctuations increased in frequency, duration, magnitude, and number of distinct levels at 53 °C. At 55 °C, we observed frequent transitions between several well-defined transmission levels. Among these levels, the original level ascribed to folded *apo*-CaM (orange dotted lines in Fig. 2) was not populated, the second level of reduced transmission (dark green) was rarely populated, the third level of reduced transmission (light green) was most probable, the fourth level of reduced transmission (blue) was next probable, and the fifth level (magenta) had low probability. Previous ensemble measurements have shown that the C-domain of *apo*-CaM unfolds between 42.5 and 47 °C, the N-domain unfolds between 55 and 58 °C,[40,42,43,45] while the melting temperature of the linker region between the two domains is unknown; solution NMR studies reveal that this flexible linker adopts a random coil conformation with partial $\alpha$-helical structure.[46] Starting from natively-folded *apo*-CaM, sequential unfolding of these three domains, presumably in various combinations, could result in at least seven additional partially unfolded conformations[19,47] and hence lead to a possible total of eight different polarizabilities with different probability, consistent with the five levels of transmission we observed. Figure 2a shows that raising the temperature to 62 °C introduced a sixth transmission level and increased the probability of observing the third and fourth level of reduced transmission. This elevated temperature also resulted in more frequent fluctuations of transmission with increased amplitude and number of different levels that were not as well-defined, suggesting that the protein conformation increased in variability as unfolding progressed. The distinct and well-defined transmission levels we observed, in particular at 55 °C, are strikingly similar to the six distinct levels of unfolding observed by Stigler et al. by force spectroscopy measurements on single CaM molecules.[19,47] In comparison, a trapped *holo*-CaM subjected to similar stepwise increases in temperature exhibited little to no increase in fluctuation up to 62 °C (Fig. 2b).

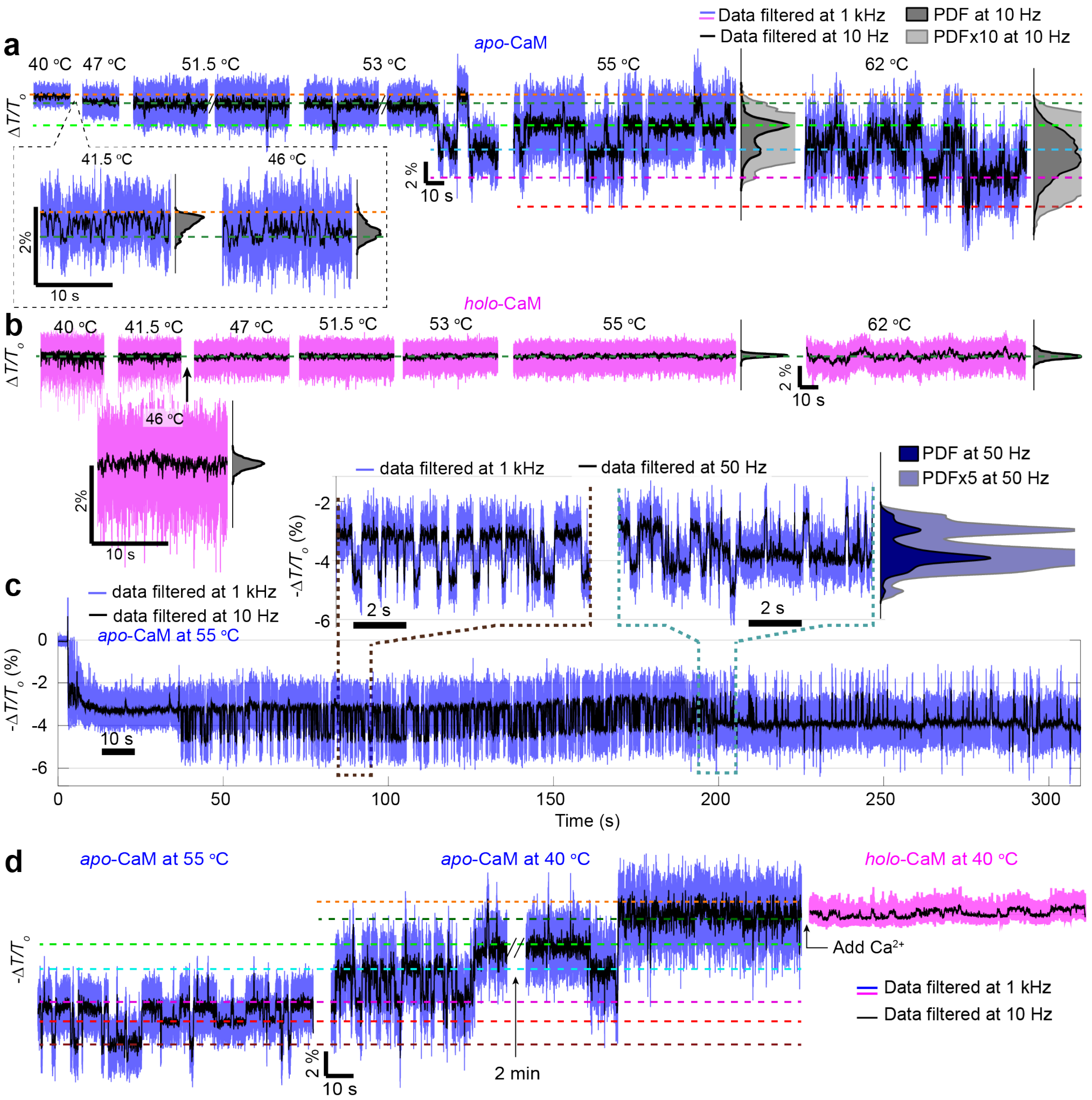

**Figure 2.** Thermal unfolding and folding of single calmodulin proteins. **a**, **b**, Transmission through a DNH with either a trapped *apo*-CaM **a**, or a *holo*-CaM **b**, in response to stepwise increases in temperature that resulted from incremental increases in laser power (see Supplementary Information SI-8). Probability density functions (PDFs) of the transmitted light through the DNH reveal six transmission levels. **c,** Continuous time trace of transmission through a DNH after trapping-induced transfer a single *apo*-CaM molecule from a solution with a temperature of ~21° C to the hotspot with a local temperature of 55 °C. **d**, Transmission recording showing the refolding process of a single *apo*-CaM molecule trapped at a temperature of 55 °C for more than 10 min before a single step decrease of the temperature of the hotspot to 40 °C. The magenta trace on the right shows reduced fluctuations in transmission after introducing 100 μM $CaCl_2$. The colored dashed lines indicate different transmission levels.

In addition to the stepwise increase in small temperature increments (Fig. 2a), Figure 2c shows a continuous time course of unfolding of a single *apo*-CaM molecule after a one-step, rapid temperature jump by trapping-induced transfer of the protein in the recording chamber with a temperature of approximately 21 °C to the plasmonic hotspot with a temperature of 55 °C. The initial and large change in transmission is the result of trapping the protein. This change in transmission is followed by a period of approximately 2 s of fluctuations and finally by further stepwise changes in transmission resulting in at least five distinct transmission levels, again similar to the results reported by Stigler et al.[19,47] The transmission levels observed approximately 200 s after trapping *apo*-CaM resemble the levels at 55 °C in Figure 2a, confirming a temperature-dependent, reproducible process of domain unfolding.[48]

Under appropriate conditions, calmodulin undergoes multiple reversible unfolding and refolding cycles in response to temperature jumps.[48] Holding a single *apo*-CaM protein trapped in the DNH at a temperature of 55 °C with frequent fluctuations between transmission levels followed by rapid reduction of the temperature in a single step to 40 °C (we determined that 90% of the total temperature change occurs within 70 μs), resulted in a step-wise return to the transmission level of natively folded *apo*-CaM within a few minutes (Figure 2d). The resulting refolded *apo*-CaM protein was capable of binding calcium ions as indicated by a strong reduction in the noise of the observed transmission upon calcium addition (Figure 2d). This result is consistent with molecular dynamics simulations that predict *holo*-CaM to be significantly less flexible than *apo*-CaM.[49]

As unfolding of a protein typically elongates its conformation,[19] the reduction in transmission upon heating *apo*-CaM in the plasmonic hotspot is consistent with an elongation-induced increase in polarizability;[28] this reduction in transmission is absent in the case of thermally stable *holo*-CaM.[44] The results in Figure 2 provide strong evidence that the observed changes in transmission are caused by changes in conformation of various domains of CaM during thermally induced unfolding or refolding. To the best of our knowledge, Figure 2 provides the first thermal unfolding trajectory of a single unmodified protein; no fluorescent labels, tethers, or other modifications are needed for this experiment. Moreover, the relatively close agreement between the observed unfolding temperatures in Figure 2 and the range of reported melting temperatures of *apo*-CaM (42.5 – 58 °C)[40,41,43] suggests that the electric field in the hotspot and the resulting gradient forces, on average, do not exert a strong effect on the observed melting temperature and hence on the conformation of proteins.

The results in Figure 2 confirm that the approach used here makes it possible to monitor conformational changes of single trapped proteins. Moreover, conformational changes of individual protein domains result in readily detectable changes in transmission, which can be monitored for more than two hours (see Figure S9) without possible limitations by photobleaching or by contamination of scanning probe tips for force spectroscopy. Based on these beneficial characteristics, we hypothesized that this approach might make it possible to monitor catalytic cycles of individual enzymes if one or several reaction steps are associated with conformational changes. To test this hypothesis, we trapped adenylate kinase (AdK) an enzyme that regulates the adenosine triphosphate (ATP) levels in cells by catalyzing the conversion of two adenosine diphosphates (ADP) to one ATP and one adenosine monophosphate (AMP).[50–53] The reaction is reversible such that it either produces ATP and AMP or two molecules of ADP.[50–52] Binding of ATP (or ADP in case of the reverse reaction) to the pocket formed between the LID and CORE domain results in a significant conformational change that closes the LID, while binding of AMP (or ADP) to the pocket formed between the NMP and CORE domain results in another significant conformational change that closes the NMP domain.[51,54] As AdK binds one ADP and then another, its conformation shifts from open, to partially closed, to fully closed.[51,55]

Figure 3a shows that in the absence of substrate, we observed only one, stable level of transmission from a single AdK protein in the plasmonic trap. Very rarely, we observed short-lived fluctuations from this level to a second level of increased transmission. After exposing trapped AdK to a buffer containing 1 μM ADP we observed transient increases in transmission levels, most frequently to an intermediate level and occasionally to a third level with the highest transmission. Increasing the substrate concentration further resulted in more frequent fluctuations and in well-defined occupancy of the third and highest level of transmission. The addition of the inhibitor Ap5A stabilized this highest level of transmission, while the presence of the inhibitor Ap6A stabilized the intermediate (second) level of transmission (Fig. 3a). Previous work has shown that AdK adopts predominantly its open conformation in the absence of substrates,[52,54,56] while the probability of occupying the closed state increases with increasing substrate concentration.[56] In addition, Pelz et al. have shown that the inhibitor Ap5A favors the closed conformation[54] while the inhibitor Ap6A favors an intermediate, partially closed conformation.[54] With this information, assignment of transmission levels observed with trapped AdK is straightforward: The first and lowest level of transmission, observed in the absence of substrate, corresponds to the open conformation. In this elongated conformation the polarizability of the protein is largest and therefore transmission is lowest as observed in Fig. 3a. The second level of transmission is occupied only transiently at any substrate concentration but it is favored in the presence of Ap6A. The six phosphate groups in Ap6A bind across both binding pockets of AdK and inhibit the closed state while stabilizing a partially closed conformation.[54] In this intermediate conformation, the polarizability of AdK is smaller than in the open conformation and larger than in the closed conformation, resulting in an intermediate level of transmission. The highest level of transmission corresponds to the closed conformation with the smallest polarizability of AdK; this conformation is favored at high substrate concentration[53,56] and in the presence of the inhibitor Ap5A (Fig. 3a).[50,54,56]

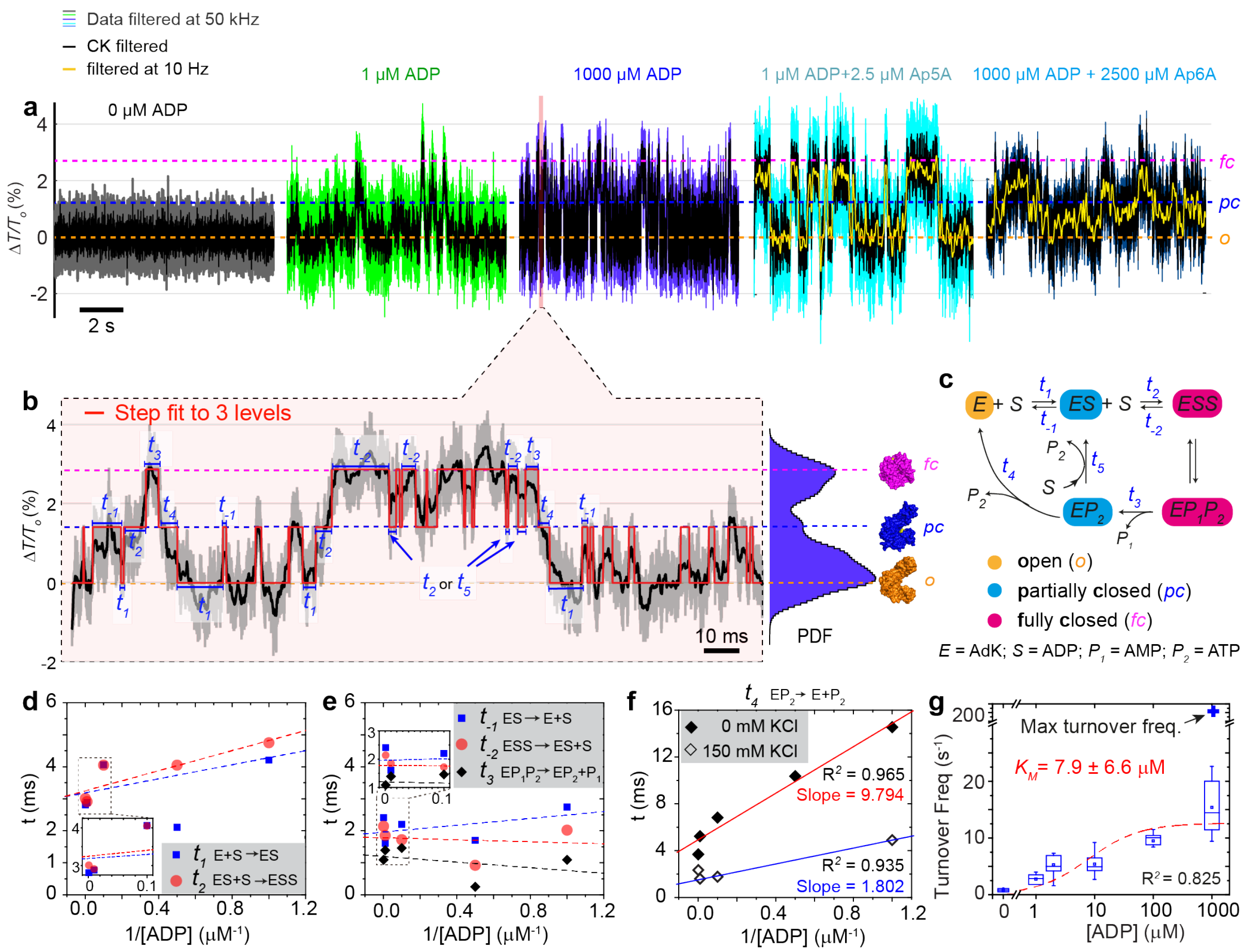

**Figure 3.** Conformational dynamics of single adenylate kinase (AdK) as a function of substrate concentration, and in the presence or absence of inhibitors. **a**, Time traces of transmission signal with a single AdK protein trapped in a DNH with varying concentrations of substrate ADP, and in the presence of inhibitors Ap5A or Ap6A. The magenta, blue and orange dashed lines represent the levels of the three conformational states corresponding to the open (*o*, orange), partially closed (*pc*, blue) and fully closed (*fc*, magenta) conformations of AdK. **b**, Illustration of a step fit to 3 levels (red) of a transmission trace (grey) after applying a CK filter (black), along with the PDF of the CK-filtered transmission trace shown on the right. The crystal structures of the conformational states are shown on the right: fully closed (1AKE), partially closed (2AK3), and open (4AKE). Here $t_1$, $t_{-1}$, $t_2$, $t_{-2}$, $t_3$, $t_4$ and $t_5$ represent the dwell time of different AdK states upon binding or unbinding substrate or products (see panel **c**, and Supplementary Information SI-11 for details). **c**, Schematic representation of the binding kinetics of AdK as it transitions between its various conformations. **d**, Dwell time before AdK binds a first ADP (blue squares) and second ADP substrate molecule (red circles) increases with decreasing ADP concentration. **e**, AdK unbinding of the first ADP, second ADP and the first product (AMP) are processes that are independent of ADP concentration. **f**, Time for AdK to release the second product (ATP) is directly proportional to the inverse of ADP concentration. This trend is stronger at low ionic strength (0 mM KCl, closed diamonds) compared to higher ionic strength (150 mM KCl, open diamonds). **g**, Turnover frequency of AdK determined by step fits to 3 levels as a function of different ADP concentrations. The $K_M$ value was obtained by fitting the median turnover frequencies to the Michaelis-Menten equation (red dashed curve). The boxes span the interquartile range, the squares are the mean values, and the horizontal lines are the median values. The blue "+" indicates the maximum turnover rate that we observed assuming a productive enzymatic cycle between partially closed and fully closed conformations without returning to the open conformation.

Assignment of these transmission levels (Fig. 3b) together with the simple model shown in Figure 3c makes it possible to follow the catalytic cycles of a single unmodified enzyme (see Figure S10, SI-11 for details) and reveals the durations ($t_1$, $t_2$, etc) of all reaction steps indicated in the cycle in Fig. 3c. Moreover, the data in Figure 3b reveals a heretofore hidden catalytic pathway of AdK. Specifically, the reaction times determined for binding the first and second ADP substrate molecule, $t_1$ and $t_2$, are directly proportional to the inverse of the ADP concentration (Fig. 3d), as expected for a bimolecular reaction (see Supplementary Information SI-11).[55] In addition, unbinding of the first ($t_{-1}$) and second ($t_{-2}$) ADP substrate molecule as well as release of the first product (typically AMP[51]), $t_3$, are independent of ADP concentration (Fig. 3e), as expected.[51] Surprisingly, however, Fig 3f shows that the release of the second product (typically ATP[51]) also depends linearly on the inverse of the ADP substrate concentration and this dependence is stronger at lower ionic strength. Why should the off-rate ($1/t_4$) of product depend on the concentration of the substrate? Considering that ATP release is the rate-limiting step[50,51] and that electrostatic interactions of $ADP^{3-}$, $ATP^{4-}$, and $AMP^{2-}$ with positively charged amino acid side chains in AdK's two active sites are important,[57] this substrate dependence of product release uncovers a previously unknow catalytic pathway. Essentially, the data in Figure 3b and 3f suggests that after the first product (AMP) is released from the active site, release of the second product $ATP^{4-}$ from the active site of AdK is accelerated by nearby binding of $ADP^{3-}$ to the free AMP site of the enzyme. In particular at high ADP concentration, it is plausible that after AMP leaves the active site, the next ADP substrate molecule may bind to this site *before* the rate-limiting release of ATP occurs. The dependence of ATP release and hence of $t_4$ on ADP concentration, shown in Fig 3f, suggests that prior binding of ADP to the AMP site favors unbinding of ATP, presumably by the additional electrostatic repulsion that $ATP^{4-}$ experiences when $ADP^{3-}$ is bound nearby at the AMP site compared to $AMP^{2-}$ bound in the AMP site or compared to an unoccupied AMP site. The observation that low ionic strength increases the ADP-dependence of unbinding of ATP supports this hypothesis of an electrostatic effect. This insight suggests that with increasing concentration of the substrate ADP, AdK can sustain productive enzymatic cycles without passing through the open state, while current understanding assumes that each catalytic cycle of AdK passes through an open state.[51,53] Yet, the single molecule data of transmission versus time from trapped AdK in the presence of increasing ADP concentrations in Figure 3b provide direct evidence for the importance of the novel pathway proposed here. For instance, Figure 3b shows that the enzyme alternates for extended periods between the highest transmission level (closed conformation) and the intermediate transmission level (partially closed conformation) at a maximum frequency of 200 $s^{-1}$, while switching all the way between the highest transmission level (closed) and the lowest transmission level (open) too rarely ($\leq 20$ $s^{-1}$) to rationalize the high previously reported turnover rates of 116 to 210 $s^{-1}$ under these $v_{max}$ conditions.[51,58] Additional insights from these single molecule transmission recordings with trapped AdK include: we never observed a single step change from the highest transmission level (closed state) to the lowest transmission level (open state), including at low ADP concentrations, indicating that the products AMP and ATP are never released simultaneously but rather sequentially. While this result may have been expected given that ATP release is the rate-limiting step[51] and that the ATP lid and AMP lid of AdK open and close independently,[51,55] we emphasize that the single enzyme recordings introduced here provide direct evidence for this sequential reaction scheme (see Supplementary Information, Figure S10b). Moreover, the rates of binding of the first ($1/t_1$) and second ADP substrate molecule ($1/t_2$) and their dependence on ADP concentration are similar, indicating no significant difference in the association rate of ADP with either active site. This insight directly rules out an allosteric mechanism. Finally, the off-rate of the second ADP ($1/t_{-2}$) is faster than that of the first ($1/t_{-1}$) as expected, since release from either of the two occupied sites leads to a return to an intermediate transmission level from the highest transmission level, while only the release from a single occupied site can cause the transition from intermediate to lowest level of transmission.

Figure 3g shows that the ability to monitor these conformational transitions during the enzymatic cycle of AdK makes it possible to determine the turnover rates as a function of substrate concentration directly from single molecule data.[51,58] A fit of the Michaelis Menten equation to this data reveals a $K_M$ value for ADP of 55 μM, in good agreement with the range of 2 - 290 μM determined by other single molecule approaches.[51,54]

To demonstrate the potential of the nanoplasmonic optical tweezer for revealing insights into conformational dynamics of native proteins that have so far not been interrogated on the single molecule level, we chose to examine citrate synthase (98 kDa).[59,60] This central enzyme of the citric acid cycle catalyzes the condensation of oxalacetate (OAA) and acetyl-coenzyme A (AcCoA) to produce citrate and coenzyme A.[59] Citrate synthase is composed of two identical subunits with one active site per subunit.[59,60] Crystallography in the absence of substrates reveals the two subunits in an open conformation, while crystals of citrate synthase in the presence of its substrates OAA, AcCoA or both reveal the subunits in a closed conformation.[60] Crystals in the presence of various transition state analogs of the three reaction steps during catalysis also reveal closed conformations.[60] Finally, exposing citrate synthase crystals in their open conformation cracks the crystals, indicating significant conformational change.[61] Based on these and other biochemical observations,[59,60] current understanding suggests the following conformations during the enzymatic cycle of dimeric citrate synthase in physiological solution: In the absence of bound substrates, the enzyme populates predominantly an open[60] and somewhat flexible[62] conformation. In contrast, binding of OAA, AcCoA or both substrates induces a closed and rigid conformation.[60,62] Substrate binding follows an ordered mechanism: OAA binds first, followed by AcCoA. Citrate synthase is substrate inhibited by AcCoA and product inhibited by citrate.[63] The three catalytic reaction steps occur presumably in the closed conformation of the enzyme with only minor changes of side chains and ligand conformation.[64] The subunits of the enzyme open to release products and to bind new substrate molecules for the next catalytic cycle.[60] Considering the two identical subunits of the enzyme,[60] the simplest possible model for its conformational states during the enzymatic cycle therefore predicts three distinct conformations: i) open-open, ii) closed-open (or open-closed), and iii) closed-closed. Early reports proposed a cooperative allosteric mechanism of the two subunits;[61] however, current understanding considers the two active sites to act independently from each other.[60] Direct evidence for various conformations of the enzyme during its dynamic activity under physiological conditions in aqueous solution is mostly unavailable, in part because ensemble assays average over all populated conformations.

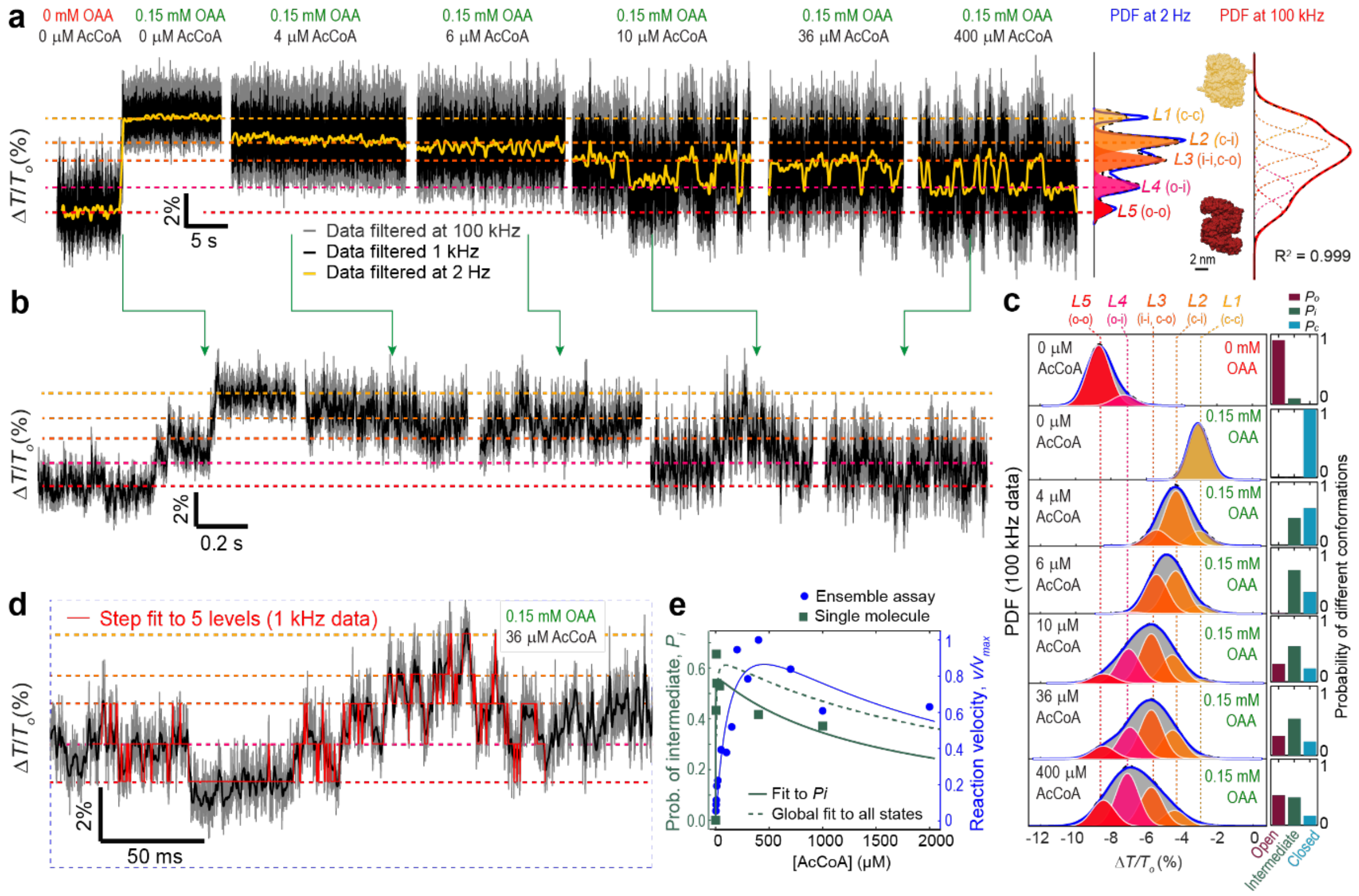

**Figure 4.** Conformational dynamics of single Citrate Synthase (CS) enzymes as a function of substrate concentration. **a**, Transmission traces of a single CS protein trapped in the DNH first in the absence of substrates and then in the presence of a constant OAA concentration and increasing AcCoA concentrations. The probability density functions (PDFs) on the right were calculated from all transmission traces after low-pass filtering at 2 Hz and 100 kHz. Dashed lines in five different colors are centered at the peaks of the 2 Hz PDF, representing five conformational states of dimeric CS: *L1* representing the closed-closed (c-c) conformation (crystal structure 4CSC), *L2* representing the closed-intermediate (c-i) conformation, *L3* representing the intermediate-intermediate (i-i) or closed-open (c-o) conformations, *L4* representing the open-intermediate (o-i) conformation, and *L5* representing the open-open (o-o) conformation (crystal structure 5CSC). The PDF calculated at 100 kHz was fitted to a function composed of 5 equidistant (± 15%) Gaussian peaks with shared width (σ ± 10%). **b**, Magnified transmission traces from panel **a** illustrating the dynamics of conformational changes. **c**, Change in the occupancy of transmission levels *L1* to *L5* as a function of substrate concentration and corresponding probabilities of the open state $P_o$, intermediate state $P_i$, and closed state $P_c$ of each subunit of CS (see Table S2 for details). **d**, Illustration of a step fit to 5 levels to a magnified section of the transmission trace recorded in the presence of 36 μM AcCoA (see Supplementary Information SI-13.1 and SI-13.2 for the analysis of set-fitted data). **e**, Comparison of the probability of the intermediate state $P_i$ of CS from single molecule data (green) with the normalized initial velocity of CS from an ensemble assay (blue) as a function of substrate concentration; both curves indicate substrate inhibition of CS by AcCoA and that $P_i$ is related to enzyme activity as predicted in the Supplementary Information SI-13.3 and SI-13.4.

Figure 4 shows that the single molecule studies of the catalytic cycle of citrate synthase presented here provide heretofore inaccessible direct evidence for at least four previously observed or postulated conformational changes. First, binding of OAA to a single unligated citrate synthase molecule in the plasmonic trap induces a significant change in transmission that occurs in two major steps (Fig. 4a). The transition from the lowest transmission level to the next occurred within 23 ms and the transition from this level to the highest transmission level occurred within 30 ms, while the overall transmission occurred within 260 ms (we confirmed that the temporal resolution of this experiment was at least as fast as 40 μs, see Supplementary Information SI-14). This change in transmission is consistent with a change from an open and elongated conformation with high polarizability in the absence of substrates[61] to a closed and compact conformation with low polarizability in the presence of OAA (see crystal structures in Fig. 4a).[60] This result provides direct evidence for an induced fit mechanism of substrate binding under physiological conditions in aqueous solution on a single-molecule level and is consistent with the strong response of citrate synthase crystals that crack upon exposure to OAA.[60] Second, trapped citrate synthase in the unligated, open conformation occasionally shows fluctuations in transmission (Fig. 4b and 4c) that are consistent with conformational variability of the open state as proposed by Roccatano et al.[62] Third, the OAA-induced closed conformation resulted in an increased but stable transmission signal without fluctuations (Figure 4a,b,c and Table S2 and S3), consistent with a stable closed-closed conformation as reported previously.[60] Fourth, when keeping the OAA concentration constant at 150 μM, increasing concentrations of AcCoA result in increasingly frequent fluctuations from a conformation with the lowest polarizability that is consistent with the closed-closed conformation of the enzyme to conformations with increased polarizability that are consistent with at least one subunit that is not fully closed (Figure 4a). These results support previous suggestions that opening of subunits of citrate synthase is required for substrate entry and product release.[60] Moreover, the results in Figure 4c support previous reports that the closed state is required for catalysis[60] since we frequently observed transmission levels that are consistent with at least one subunit in the closed state (*L2*) even at substrate concentrations above 36 μM, and hence above the $K_M$ value of AcCoA (5 to 8 μM),[59,60,65]

Figure 4 also shows that at least six aspects of the current understanding of the conformational changes of citrate synthase during its catalytic cycle are too simple, incomplete, and in some instances, incorrect. The first new insight is that during enzymatic activity, dimeric citrate synthase populates more than the three conformations predicted by the currently used model.[60] The two distributions of transmission levels (PDFs) in Figure 4a show for instance that increasing the concentration of AcCoA beyond its $K_M$ value results in the appearance of at least five levels indicating at least five distinguishable enzyme conformations. This distribution of transmission levels can be fit very well with five Gaussian peaks that are centered on approximately equidistant levels and that each have the width (i.e., standard deviation σ) of the peak that represents transmission level *L1* of the closed-closed state. Based on this observation, we considered the simplest possible model that predicts five distinguishable conformations of a dimeric enzyme. This model assumes each subunit to occupy one additional, intermediate conformation, that is partly open and lies between the closed and open state. Since we observed five transmission levels that appeared approximately equidistant from each other, we assumed that the change in transmission when one subunit transitions from the closed to the partially open state is half the change observed when one subunit transitions from the closed state to the fully open state (not making this assumption would lead to six conformations and six transmission levels with non-uniform distances between them). As summarized in Table 1, this model of a single trapped citrate synthase predicts the five distinguishable, conformation-induced transmission levels *L1* to *L5* along with the corresponding conformations of subunit 1 and 2 of citrate synthase. To assign transmission levels to conformations, we took advantage of the two extreme transmission levels: level *L1* observed in the presence of OAA only, which corresponds to the closed-closed conformation and level *L5* observed in the absence of substrates, which corresponds to the open-open conformation.[60]

Table 1. Assignment of transmission levels to subunit conformations of citrate synthase.

| Trans-mission level | Corresponding conformations of subunit *1* and *2* | Relative occupancy of transmission levels and corresponding probability of subunit conformations | Probability of a subunit to be in the closed $P_c$, intermediate $P_i$, or open $P_o$ conformation.[a] |
|---|---|---|---|
| *L1* | *1c2c* | $O_{L1} = P_{1c} \times P_{2c}$ | $P_c = \sqrt{O_{L1}}$ |
| *L2* | *1c2i and 1i2c* | $O_{L2} = P_{1c} \times P_{2i} + P_{1i} \times P_{2c}$ | |
| *L3* | *1i2i and 1c2o and 1o2c* | $O_{L3} = P_{1i} \times P_{2i} + P_{1c} \times P_{2o} + P_{1o} \times P_{2c}$ | $P_i = \sqrt{O_{L3} - 2 \times \sqrt{O_{L1}} \times \sqrt{O_{L5}}}$ |
| *L4* | *1i2o and 1o2i* | $O_{L4} = P_{1i} \times P_{2o} + P_{1o} \times P_{2i}$ | |
| *L5* | *1o2o* | $O_{L5} = P_{1o} \times P_{2o}$ | $P_o = \sqrt{O_{L5}}$ |

[a]In order to relate the occupancies of transmission levels 1 to 5 ($O_{L1}$ to $O_{L5}$) to the probability that a subunit is in the closed ($P_c$), intermediate ($P_i$) or open ($P_o$) conformation, we assumed that the probability of subunit 1 to be open ($P_{1o}$) is the same as the probability of subunit 2 to be open ($P_{2o}$) because both subunits are identical, hence $P_{1o} = P_{2o} = P_o$. The same applies to the probability of the intermediate conformation and the closed conformations of both subunits, hence $P_{1i} = P_{2i} = P_i$ and $P_{1c} = P_{2c} = P_c$. Finally, we assumed that the probability of subunit 1 to assume any one of the three conformations is independent of the probability of subunit 2 to assume any one of the three conformations, hence the probability for conformation *1c2c* is $P_{1c} \times P_{2c} = P_c \times P_c = P_c^2$, the probability for conformation *1c2i* is $P_{1c} \times P_{2i} = P_c \times P_i$, etc.

With this assignment of conformations, the relative occupancies of each transmission level $O_{L1}$ to $O_{L5}$ from the five-peak fit to their distribution (Fig. 4c) makes it possible to estimate the probability of the closed ($P_c$), intermediate ($P_i$), or open state ($P_o$) of each subunit as a function of substrate concentration (see all the values in Table S2). Figure 4c shows the probabilities of these three conformations for each AcCoA concentration and illustrates that the intermediate, partially open level introduced here is the most probable state of the subunits of enzymatically active citrate synthase, while the open state is not populated below AcCoA concentrations of 10 µM. Thus, the second new insight is that during enzyme activity, the subunits of citrate synthase frequently populate an intermediate, partially open conformation and that this conformation is more probable than the closed or open conformation. Consequently, the third new insight is that complete opening of each subunit of citrate synthase may not be required during the enzymatic cycle in physiologic solution; instead, partial opening to the intermediate conformation appears to be sufficient for product release and substrate binding. The fourth new insight is that the single protein data presented here directly rules out an allosteric mechanism for the two subunits of citrate synthase. Allostery would favor the three symmetric conformations closed-closed (corresponding to *L1*), intermediate-intermediate (*L3*), and open-open (*L5*) at the expense of the hybrid conformations closed-intermediate (*L2*) or intermediate-open (*L4*). Figure 4c, however, indicates no preference for the symmetric conformations at any concentration of AcCoA above 0 µM. Instead, we recorded fluctuations between adjacent transmission levels rather than fluctuations between every other transmission level (see Supplementary Information SI-13.3, Figure S12). The fifth new insight is that as the concentration of AcCoA reached at least 36 µM, the recorded transmission levels of trapped citrate synthase represented mostly the intermediate levels *L2, L3,* and *L4*, while occasionally and transiently reaching the two extreme transmission levels *L1* or *L5*. This result confirms that citrate synthase indeed samples the entire range of conformations from the closed-closed to the open-open conformation during its dynamic activity in aqueous solution. It appears, however, that the fully open conformation is rare and only populated at high AcCoA concentration when substrate inhibition by AcCoA occurs. The sixth new insight is that the probability of the intermediate state $P_i$ is proportional to the activity of citrate synthase (see Supporting Information SI-13.4) and captures the previously reported substrate inhibition by AcCoA (Figure 4e). This dependency makes it possible to estimate an apparent $K_M$ value from the single molecule data resulting in a value of 0.4 to 4.4 μM (see Supplementary SI-13.4 for details). This estimate approaches the range of previously published $K_M$ values of 5 to 8 μM from ensemble assays[59,60,65] and shows that the approach introduced here quantifies enzyme activity without the need for chromogenic or otherwise detectable substrates or products. Instead, the signal can be directly inferred from activity-dependent conformational changes as long as these are detectable.

**Conclusion and outlook**

The approach introduced here makes it possible to monitor the conformational changes of single unmodified proteins during allosteric change, thermal unfolding and catalytic cycling of enzymes with a time resolution as fast as 40 μs and possibly faster. The results from four different proteins are consistent with the hypothesis that elongation of protein conformation increases the refractive index in the hotspot area and thereby results in a detectable decrease in the transmission of scattered light through a plasmonic DNH with a single trapped protein. This dependency enables assignment of the experimentally observed transmission levels to various protein conformations: the most open and elongated conformation of a protein corresponds to the lowest transmission level, intermediate elongations to intermediate transmission levels and the most compact conformation corresponds to the highest transmission level. This insight makes it possible to monitor thermal unfolding and refolding of domains of single proteins and well as following various conformations of a single protein through its enzymatic cycle, revealing time constants of all detectable steps as well as heretofore unknown enzymatic pathways and conformations.

The approach is particularly well suited for monitoring proteins that undergo significant conformational changes on time scales reaching from tens of microseconds to hours and it works best for large proteins since the restoring force in the plasmonic optical trap increases with particle size. The thermal unfolding and refolding experiments with *apo*-CaM showed, however, that even relatively small proteins with a molecular weight of 16.7 kDa can be trapped and interrogated for two hours (see Supplementary Information Figure S9). Compared to the two most widely used methods for interrogating single protein dynamics, smFRET and smFS, the approach introduced here has the advantage that it does not require site-directed labeling or tethering of the proteins under investigation; the method interrogates native proteins in aqueous solution without any modification. The approach circumvents technical difficulties associated with single molecule fluorescence techniques such as photobleaching, elimination of background fluorescence or distinguishing the emission dynamics of fluorophores from conformational dynamics and it circumvents challenges of smFS such as tip contamination or attachment of a single tether to a probe tip or bead and to a defined location on a protein. In terms of disadvantages compared to smFRET and smFS, the approach does not give direct information about the change in distance between two (or more) defined locations during conformational changes. Rather, the approach monitors the effect of an overall change in protein conformation on the local refractive index in the hotspot area. Another limitation is that after passivating the surfaces in the hotspot area with protein resistant coatings, the approach requires significant laser power to trap single proteins in their native state; in the experimental configuration we used, the local temperature in the hotspot area was 35 – 40° C. While this temperature range is ideal for the interrogation of proteins at human body temperature, it may be too high for thermally fragile proteins. Employing advanced plasmonic nanostructures such as optical metamaterials,[66] optimizing the device geometry of the DNH, or taking advantage of a tunable laser to optimize its wavelength with respect to the specific resonance frequency of each nanofabricated plasmonic device may increase the trap stiffness. This optimization would enable a reduction in laser power to limit heating in the hotspot area while increasing the signal to noise ratio (SNR). An improved SNR may ultimately make it possible to record conformational dynamics of single native proteins at sub-microsecond time resolution, possibly reaching the low nanosecond range if an avalanche photodiode with 1 GHz bandwidth[3,28] would be used and the conformational dynamics result in a detectable change at this bandwidth.

Future applications of the approach beyond the ones reported here may entail screening for ligands that stabilize or destabilize protein conformation,[67] monitoring the formation of single protein complexes or protein aggregates, and assessing the conformational variability of intrinsically disordered proteins as well as their change in conformation upon binding to target proteins.[68] Should the approach also be applicable to membrane proteins in detergent micelles or lipid nanodiscs,[3] then it may report conformational changes in response to activation of receptors, transporters or ion channels. More generally, we suggest that this approach provides opportunities by complementing the rapid advances of structural biology from cryo-EM with insight into conformational dynamics and transition paths of single unmodified proteins under physiologic conditions in aqueous solution.

**Acknowledgments.** This research was supported by Oxford Nanopore Technologies (M.M., Grant Number 350509-N016133), the Swiss National Science Foundation (SNSF), Grant No. 200021_169304 and Grant No. 200020_197239 to M.M., the Adolphe Merkle Foundation and the University of Fribourg. E.B.-U. was funded by the European Union's Horizon 2020 research and innovation programme under the Marie Skłodowska-Curie grant agreement No 741855. S.A. and M.M. are grateful to the Michael J. Fox Foundation for Parkinson's Research (MJFF-009813). S.A. and L.B. also acknowledges the SNSF SPARK program (grant number 195960). S.B. acknowledges support from the SNSF through the National Center of Competence in Research (NCCR) Bio-Inspired Materials (25%). R.G. acknowledges funding from Natural Sciences and Engineering Research Council of Canada (RGPIN-2017-03830). G.P.A. acknowledges financial support by the Swiss National Science Foundation (SNSF), Grant No. 200021_184687. C.Y. would like to thank Dimitri Vanhecke from the Adolphe Merkle Institute and the team from the Center for MicroNanotechnology (CMi) from the École Polytechnique Fédérale de Lausanne (EPFL) for training and technical discussions.

**Author contributions.** C.Y. and M.M. conceived the project. C.Y., E. K. and M.M. designed the experiments. C.Y., E.K. and C.F. performed all data collection. C.Y. and E.K. fabricated the DNH samples. C.Y., E.K., A.I., A.G., S.B. and M.M conducted data analysis. C.Y. and E.B.-U. performed the COMSOL and FDTD simulations. A.I. provided help with kinetic models for enzyme activity. S.A. and L.B. provided knowledge about proteins and help with protein preparation. J.L. helped with the design of the fluidic setup. G.P.A. and R.G. provided support with the optical setup. C.Y. and M.M. wrote the manuscript with input from E.K., E.B.-U., A.I., and A.G. All authors discussed the results and commented on the manuscript.

Supplementary Information

# Watching Single Unmodified Enzymes at Work

Cuifeng Ying,[1, 2] Edona Karakaçi,[1] Esteban Bermúdez-Ureña,[1,5] Alessandro Ianiro,[1] Ceri Foster,[1] Saurabh Awasthi,[1] Anirvan Guha,[1] Louise Bryan,[1] Jonathan List,[1] Sandor Balog,[1] Guillermo P. Acuna[3], Reuven Gordon[4], Michael Mayer[1,*]

[1]Adolphe Merkle Institute, University of Fribourg, Chemin des Verdiers 4, CH-1700 Fribourg, Switzerland.

[2]Advanced Optics and Photonics Lab, Department of Engineering, School of Science and Technology, Nottingham Trent University, United Kingdom

[3]Department of Physics, University of Fribourg, Chemin du Musée 3, CH-1700 Fribourg, Switzerland.

[4]Department of Electrical and Computer Engineering, University of Victoria, Victoria, British Columbia V8P 5C2, Canada

[5]Current address: Centro de Investigación en Ciencia e Ingeniería de Materiales and Escuela de Física, Universidad de Costa Rica, San José, 11501, Costa Rica

Corresponding Author Email: michael.mayer@unifr.ch

# Table of Contents

## SI-1 Materials and Methods

**Chemicals.** We purchased all the chemicals from Sigma-Aldrich except for $(Me)_2SiCl\text{-}(CH_2)_m\text{-}EG_nOCH_3$ and HS-$C_{11}$-$EG_4$-Carboxybetaine, which we obtained from ProChimia (ProChimia Surfaces, Poland).

**Fabrication of Nanostructures in Films of Gold**. We deposited a 30 nm silicon nitride ($SiN_x$) layer onto fused silica wafers (thickness 550 µm) by means of low-pressure chemical vapor deposition (LPCVD) at 800 °C. We overlaid the resulting silicon nitride layer with a 5 nm thick film of Ti and followed by a 100 nm thick film of Au, which we deposited by means of electron-beam evaporation (Leybold Optics LAB 600H) with a substrate temperature of 190 °C. The wafers were diced into 10 mm × 10 mm chips (Disco DAD321). We used a focused ion-beam (FIB, Thermo Scientific Scios 2 Dual Beam) with a Gallium ion source operated at 30 keV and a current of 1.5 pA to fabricate the double nanohole (DNH) structures in the gold film.[1] We imaged the DNH structures using scanning electron microscopy (SEM) in top-view and tilted modes. All DNH structures used for experiments with proteins had gap sizes in the range of 10 to 25 nm.

**Surface Passivation.** To passivate both the gold and $SiN_x$ surfaces, we immersed the samples in a solution containing 2 mM $Me_2SiCl$-$C_{11}$-$EG_6$OMe and 2 mM Poly(ethylene glycol) methyl ether thiol (PEG thiol, average molecular weight of 800 g/mol) in ethanol or 2 mM $Me_2SiCl$-$C_{11}$-$EG_6$OMe and 2 mM HS-$C_{11}$-$EG_4$-Carboxybetaine in ethanol directly after FIB fabrication. To minimize oxidation of the gold sample before surface passivation, we freshly prepared these solutions immediately before each use in a glove box filled with nitrogen and also performed the surface passivation in the glove box. After an over-night incubation, we rinsed the samples thoroughly with ethanol, dried them under a jet of $N_2$ and stored them dry at 4 °C until use.

**Preparation of Protein Solutions.** We purchased all proteins, including Ak1 human, (adenylate kinase, SRP6121), hemoglobin human (H7379), Calmodulin bovine (CaM, C4874) and Citrate Synthase from porcine heart (CS, C3260) from Sigma-Aldrich, Switzerland.
*Adenylate kinase.* The buffer used for all AdK experiments was an aqueous solution containing 10% (v/v) glycerol, 2 mM $MgSO_4$, and 20 mM Tris-HCl at pH 7.5. All substrates and inhibitors used in these experiments were dissolved in this buffer. We used 1.1 µM AdK in this buffer to trap the enzyme.
*Calmodulin.* The $Ca^{2+}$-free buffer contained 10 mM HEPES, 0.1 M KCl, and 1 mM EGTA at pH 7.4. The $Ca^{2+}$-rich buffer contained 10 mM HEPES, 0.1 M KCl, 1 mM $MgCl_2$, and 0.1 mM $CaCl_2$ at pH 7.4.[2] We used 10 µM calmodulin in $Ca^{2+}$-free buffer for the trapping experiments.
*Hemoglobin.* We prepared 10 µM methemoglobin (Met-Hb) in 0.1 M phosphate buffer (PB, pH 7.4). To deoxygenate the solution and to reduce the $Fe^{3+}$ in the heme group of methemoglobin to $Fe^{2+}$ such that we obtained hemoglobin in its deoxygenated state, we added sodium hydrosulfite ($Na_2S_2O_4$) to the solution at a $Na_2S_2O_4$:Hb ratio of 200:1.[3] To prepare the oxygen-rich buffer, we bubbled compressed air into the PB buffer (without $Na_2S_2O_4$) for 20 min before injecting the solution into the trapping chamber. We measured the spectra of the Hb solutions using an Analytik Jena Specord 50 PLUS spectrometer with a path length of 10 mm (see the spectra in Figure S6). The deoxy-Hb solutions used in the trapping experiments contained 10 µM Hb in deoxygenated buffer (2 mM $Na_2S_2O_4$ in BP buffer).
*Citrate synthase.* The buffer used for citrate synthase (CS) experiments contained 100 mM KCl, 50 mM Tris, and 0.3 mM 5,5-dithio-bis-(2-nitrobenzoic acid) (DTNB) at pH 8.0. We used 1 µM CS in this buffer for trapping experiments.

*Enzymatic assay for citrate synthase.* We performed the ensemble assay of enzyme activity by following a protocol by Srere et al with small adjustments.[4] We prepared two stock solutions: solution A contained KCl, Tris-HCl, oxalacetate (OAA) and acetyl-coenzyme A (AcCoA) at pH 8.0, and solution B contained citrate synthase and DTNB. We mixed solution A and B so that the mixture contained 0.24 nM CS, 0.15 mM OAA, 100 mM KCl, 50 mM Tris, 0.3 mM DTNB and various desired AcCoA concentrations. We monitored the absorbance (at 412 nm) of the solutions immediately after mixing over time using an Analytik Jena Specord 50 PLUS spectrometer with a path length of 10 mm (PMMA semi-micro cuvettes, BRAND,759115).

**Nanoparticles.** We used polystyrene beads with a size distribution of 22 ± 2 nm (Thermo Scientific, 3020A) and silica beads with a size distribution of 21.6 ± 2.1 nm (nanoComposix, SISN20-25M) for the experiments that trapped nanoparticles. In both trapping experiments, we used the same DNH structure with a gap of 40 nm and the beads were dispersed in deionized water to a final concentration of 0.1 mg/ml.

**Fluidic System.** We printed flow cells with a 3D printer with a resolution of 50 µm, using a Form 2 printer with Clear V4 resin (Formlabs Inc., USA). We sealed the samples in the flow cell with a cover glass (thickness 0.17 mm) by using a two-component silicone-glue (Twinsil, Picodent, Germany). Double-sided tape (ARcare92712, Adhesive Research, Inc.) separated the cover slide and the DNH sample, forming a liquid channel with a height of 50 µm and a volume of 3.5 µL in the flow cell. A syringe pump (Harvard Apparatus, US) controlled the flow rate and flow direction through a 6-port valve. After trapping a protein, we perfused buffer through the recording chamber at a flow rate of either 0.01 mL/min or 0.005 mL/min to obtain buffer exchange. To ensure complete buffer exchange, we injected at least 300% of the system's volume, including that of the tubing from the valve to the trapping chamber, for solution exchange.

**Optical Setup.** We purchased all optical components from Thorlabs. The optical trapping system was a modification of an optical tweezer kit (OTKB/M). A polarizer and a half-wave plate adjusted the polarization of an 852 nm laser (FPL852S) such that the polarization was perpendicular to the line connecting the center of the two nanoholes that form the DNH. A 100× oil-immersion objective with numerical aperture of 1.25 (Nikon, Tokyo, Japan) focused the laser into a spot of roughly 1.5 µm diameter. The incident laser power before the objective was 23 mW, leading to a power density of 13 mW/$\mu m^2$ at the DNH sample. A silicon avalanche photodiode (APD120A, Thorlabs) detected the intensity of the transmitted light and converted it into a voltage signal.

**Data Acquisition.** The bandwidth of the APD (APD120A, Thorlabs) we used is 50 MHz. The fastest accessible sampling rate of our setup, determined using an oscilloscope (DPO4032, Tektronix), was 500 MS/s, more than 5-times larger than the bandwidth of the APD. The APD was hence the bandwidth-limiting component of the setup providing a maximum time resolution of 20 ns for the electronic hardware. For practical reasons (e.g., to handle files, limit file size, accelerate analysis, and reduce noise), we carried out most recordings of transmission with a data acquisition (DAQ) card. We used two DAQ cards from National Instruments: either a PXI 6221 with a sampling rate of 250 kS/s, or a PCI-5922 with a sampling rate of either 625 kS/s or 15 MS/s, leading to a best possible time resolution of the experimental recording hardware of ~12 µs (PXI 6221 at 250 kS/s), ~5 µs (PCI-5922 at 625 kS/s) or ~200 ns (PCI-5922 at 15 MS/s). The sampling rate for each experiment is specified in the main text.

**Data Analysis.** We analyzed all the data by programs coded in MATLAB (MathWorks).

*Autocorrelation Function.* We calculated the autocorrelation function by processing time traces of the raw data with a duration of 2 s using the MATLAB function **autocorr.m**. We then fitted the autocorrelation function with an exponential decay, weighting the fit to the magnitude of the autocorrelation curve to reduce the importance of data points near the tail end of the curve. We first fitted the autocorrelation curve with a single-exponential decay function to check the quality of the fit. If the majority of the curves did not fit well to a single-exponential decay function, we fit the curve with a double-exponential decay. In these cases, we attributed the first time constant to the proteins' trapping stiffness (Figure S2) and the second time constant to conformational fluctuations.

*Data filtering.* Unless otherwise stated, we used a zero-phase Gaussian low-pass filter to filter all the raw data to a desired cutoff frequency, as specified in each figure. To identify the conformational states of AdK, we filtered the transmission traces by using a Chung-Kennedy sliding window nonlinear step filter (CK filter) [5,6] with a window size of 500 points (2 ms) that is commonly used for the analysis of smFRET data.[7]

*Normalization of transmission traces.* As the transmission of a DNH varies with DNH size and laser power, it is difficult to compare the conformation of the trapped proteins between different experiments. We therefore normalized all recorded transmission traces to a baseline to remove the effect induced by size variations of the nanostructures. Unless otherwise stated, all the traces shown in this work are normalized with the function $\Delta T/T_0 = (T-T_0)/T_0$, where $T$ is the intensity of the transmitted light recorded by the APD, and $T_0$ is a baseline transmission value. This value can be either the mean of the transmission trace or the transmission intensity of the DNH with a protein in its dominant conformation, depending on the experimental conditions as specified below.

Figure 1. For the bead trapping experiments in Fig. 1e, $T_0$ is the transmission of the DNH in the absence of protein. In Fig. 1f, we normalized both traces to the transmission of the DNH in the presence of deoxy-Hb.

Figure 2. Since we varied the temperature in the hotspot for CaM unfolding and refolding experiments by taking advantage of different laser powers, we normalized the transmission traces of CaM trapping to the mean of each trace in Fig. 2a, 2b and 2d. To assign transmission levels, we aligned the transmission traces shown in Figure 2a, 2b and 2d, in such a way that the level at the end of a transmission trace aligned with the level at the beginning of the next transmission trace at the next higher or lower temperature (in addition to the transmission traces shown in Figure 4, we recorded traces at intermediate temperatures in small temperature increments to carry out this alignment). For the long trapping trace in Fig. 2c, we first removed a linear drift of the trace by using the MATLAB function - **detrend.m**. We then normalized the detrended trace to the transmission of the empty DNH. Occasionally we observed an increase in transmission upon trapping of single proteins due to variations in geometry between different DNH structures,[9] such as in the case of CaM in Fig. 2c and 2d. In these cases, we plotted $-\Delta T/T_0$ instead of $\Delta T/T_0$ to maintain visual consistency between different figures. We confirmed that in this case an increase in polarizability (for instance by an elongation of protein conformation) leads to an increase of transmission (equivalent to a decrease in $-\Delta T/T_0$).[8]

Figure 3. We normalized the transmission traces recorded with AdK trapped in the DNH to the mean of each trace. We then moved the traces such that all the traces are bottom aligned. This alignment assumes that the open conformation of AdK is observed at least sporadically in all conditions.

Figure 4. All the traces in Figure 4 are normalized to the transmission intensity of the DNH with a CS protein occupying its closed conformation. To minimize noise-related artefacts, we filtered the traces with a cut-off frequency of 10 kHz and defined $T_0$ as the maximum of each filtered trace. This normalization assumes that CS populates its closed-closed (c-c) conformation at least a few times in all conditions.

*Step fit for AdK and CS experimental data.*

Step fit to 6 levels for Hb data. The PDF calculated from the raw traces of the Hb oxygenation experiments revealed six peaks, as indicated by the pink box in Figure 1f, and therefore we performed a step-fit to six levels. We set a threshold in the middle of each pair of adjacent levels, resulting in a total of five thresholds, and assigned each data point to a level according to these five thresholds.

Step fit to 3 levels for AdK data. We calculated the PDFs of CK-filtered transmission traces for AdK experimental data and assigned the first and last peak in each PDF to the level of the open (*o*) conformation and fully closed (*fc*) conformation, respectively. We then positioned the level of the partially closed (*pc*) conformation halfway between the *fc* and *o* levels, as illustrated in Fig. 3b. We set two thresholds directly in the middle of the levels, according to which we generated a step fitted trace to 3 levels by assigning each point of CK-filtered data to one of these levels.

Step fit to 5 levels for CS data. We identified 5 levels, $L_1$ to $L_5$, in the peaks of the PDF calculated from the 2-Hz low-pass filtered traces of all conditions, as indicated by the dashed lines in Fig. 4a. We set a threshold in the middle of each pair of adjacent levels, resulting in a total of four thresholds. To reduce noise for step-fitting, we filtered the raw traces to 1 kHz and assigned each data point to a level according to these four thresholds.

**Finite-Element Simulation.** We calculated the laser-induced heating effect by finite-element simulation (FES) with COMSOL Multiphysics 5.2a (COMSOL Inc.). See SI-8 for further information.

**Finite Difference Time Domain optical simulations**

We modeled the optical properties of the double nanohole structures (DNHs) with a finite-difference time-domain-based commercial software (FDTD Solutions, Lumerical, Inc., Canada). The material stack we considered was comprised of a 100 nm Au layer (Johnson & Christy), 5 nm of Ti (Palik) and 30 nm of $SiN_x$ (refractive index (RI) of 2), on top of a glass substrate layer (RI of 1.45). The surrounding medium was considered to be water with a RI of 1.333. A top-view high resolution SEM image of a representative DNH was processed into a binary profile and imported using the image import wizard. This structure was extruded and, to act as a hole through the multilayer stack, it was assigned the same refractive index as the surrounding medium. To account for the tapering observed in the DNH structures (Fig. 1 of the main text), we introduced an 'n side truncated pyramid' structure, also with a RI of 1.333 and with parameters such that the gap width transitioned from ~18 nm at the bottom of the Au/Ti layer to ~40 nm at the top of the Au layer. The proteins were modelled as spheres with a fixed RI. When possible, depending on the location of the dielectric particles considered, symmetry boundary conditions were used for the 'in-plane' directions to reduce the computational time. The DNH structure was illuminated with a total-field scattered source (1.2 µm × 1.2 µm lateral dimensions) incident normal from the Au layer side, and with a polarization along the gap direction. To mimic closely the experimental setup, where an air objective with a numerical aperture of 0.25 was used for the light collection, the transmitted light was calculated using a far field analysis group ('far field change index') located 1.2 µm below the structure, which accounted for the limited angle collection and losses at the glass/air interface. We used a mesh size of 1 nm for the simulations shown in Figure 1. For the electric field profile (Fig. 1b), we used 2D frequency domain profile monitors placed across the gap and at the Au/Ti transition.

## SI-2 Optical properties of double-nanohole structures.

The double-nanohole (DNH) structures used throughout this work were based on those reported by Ghorbanzadeh *et al.*[10] As described in the Methods section, we used an SEM top view image of one of the DNH structures to simulate the expected optical response of the structures using the 3D-FDTD method. Figure S1a shows the optical response of this structure over a broad range of wavelengths when excited with transversally polarized light (i.e., along the gap), for the case of a non-tapered gap and accounting for a of $SiN_x$ removal depth of 15 nm during the fabrication of the structure by FIB. The blue and red dashed lines correspond to the electric field enhancement measured with a point field monitor placed at the bottom of the gap ($x$ = 8 nm, $y$ = 0, $z$ = 0 nm), and the transmission through the DNH (limited to a collection with NA = 0.25), respectively. The two peaks observed in the electric field enhancement curve correspond to the zeroth-order Fabry-Perot (FP0) mode ($\lambda \sim$ 1990 nm) and the wedge plasmon polariton (WPP) mode ($\lambda \sim$ 760 nm). As shown by Ghorbanzadeh *et al.,*[10] tapering of the DNH gap region can lead to an enhanced wedge plasmon polariton mode, which we also observe when modifying the DNH structure used in these simulations to account for an expanding width of the gap from 18 nm at bottom to 40 nm at the top of the metallic layer (solid lines in Fig. S1a). The introduction of the taper, apart from the small enhancement observed for the electric field focused at the bottom of the gap, leads to increased transmission values compared to the non-tapered DNH structure, with a noticeable peak in the transmitted light spectra at the resonance mode ($\lambda \sim$ 740 nm).

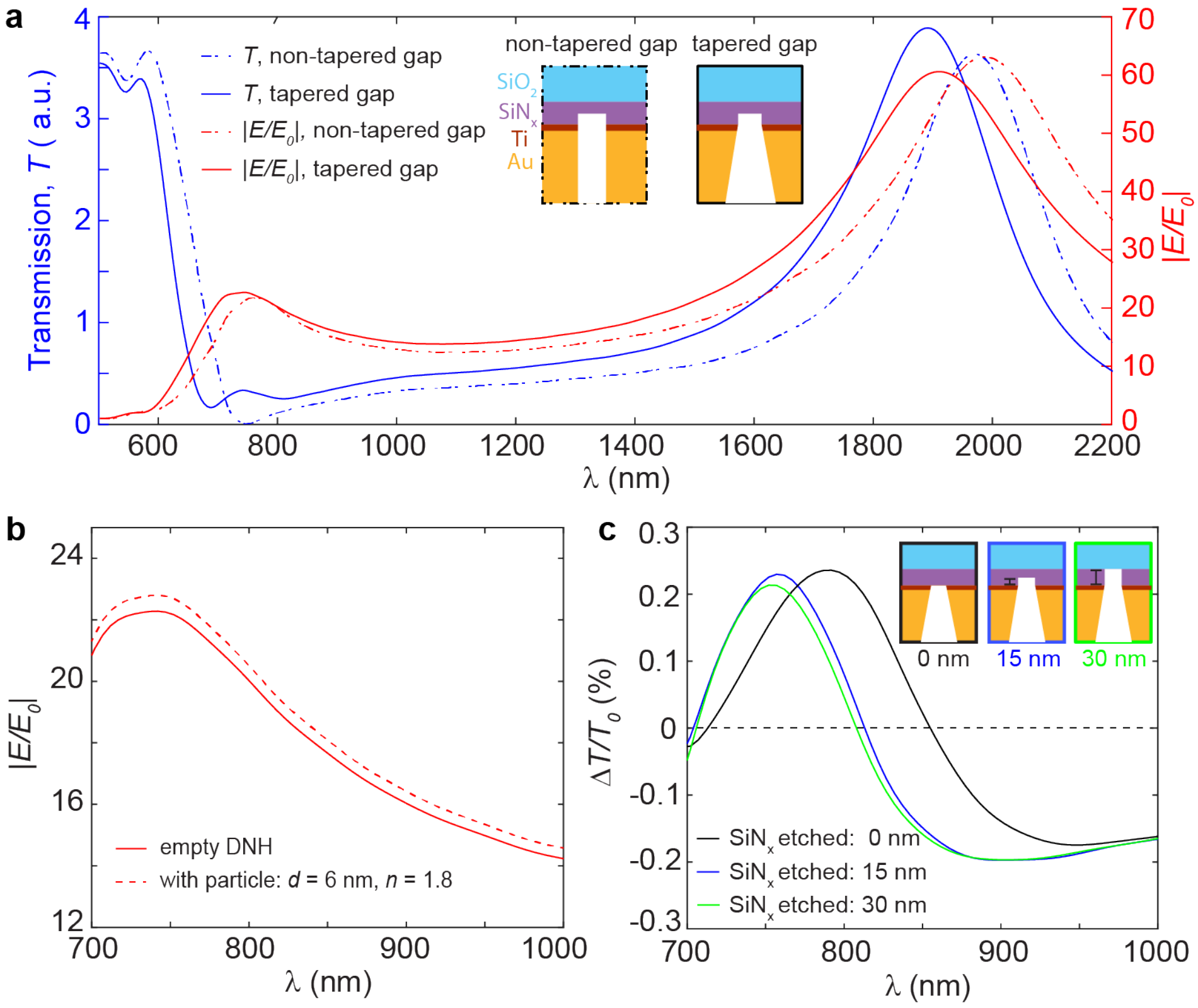

**Figure S1**. **Simulated optical properties of DNH structures. a**, Simulated transmission (blue) and electric field enhancement (red) spectra of DNH structures with a tapered gap (solid lines) and a non-tapered gap (dash-dotted lines). **b**, Simulated electric field enhancement as a function of wavelength for the empty DNH (solid curve) and in the presence of a dielectric spherical particle trapped in the hotspot (dashed curve) that has a diameter of 6 nm and a refractive index of 1.8. **c**, Effect of the depth of removal (etching) of $SiN_x$ during the FIB fabrication process on the expected change in transmission through the DNH structures in response to a dielectric particle trapped at the bottom of the gap of the gold.

The plasmonic optical trap presented here relies on the self-induced back-action (SIBA) concept, where the presence of the trapped object enhances the local electric field as first shown by Juan *et al.*[11] We used 3D-FDTD simulations by placing a point field monitor near the gap surface at the bottom of the DNH ($x$ = 8 nm, $y$ = 0 nm, $z$ = 0 nm) to record the electric field amplitude upon excitation of the DNH. Figure S1b shows the electric field enhancement as a function of wavelength in the presence or absence of a dielectric spherical particle ($d$ = 6 nm, $n$ =1.8). The electric field strength increased after introducing a particle to the hotspot regardless of whether the transmission increased or decreased with each trapping event.

We also investigated the effect of the milling depth of $SiN_x$ on the change in transmission through the DNH structures with and without a dielectric particle present at the bottom of the gap. Figure S1c shows $\Delta T/T_0$ as a function of wavelength for three typical scenarios: 1) the DNH structure ends directly at the Ti/$SiN_x$ interface (milling depth = 0 nm), and two milling depths corresponding to 2) half-way into the $SiN_x$ (depth = 15 nm), and 3) all the way to the $SiN_x$/$SiO_2$ interface (depth = 30 nm). We observed a similar behavior to that reported by Ghorbanzadeh *et al.*[10], who showed an initial shift when using the DNH etched slightly beyond the metal-dielectric interface. Notably, depending on how much the FIB process mills into the underlying dielectric layer and the operation wavelength of the experimental setup, the results in Figure S1 show that either positive or negative transmission changes can occur in response to a trapping event of a particle with a refractive index that is larger than the refractive index of the surrounding medium. Given that throughout this work we observed mainly negative transmission events upon trapping, we used an estimated milling depth of 15 nm for the simulations presented in Figure 1 of the main text. In the few trapping experiments when trapping of a protein led to an increase in transmission, we consistently observed that a change that further increased the refractive index of the particle in the trap also led to a further increase in transmission. These results are analogous to the majority of experiments in which trapping of a protein reduced the transmission and for which a change that further increased the refractive index (or polarizability) of the particle in the trap led to a further decrease in transmission as shown in Figure 1.

## SI-3 Trapping stiffness for trapped proteins of different sizes.

The trapping stiffness $\kappa$ (N/m) of an optical trap can be estimated by Equation S3.1:[9,12–14]:

$$\kappa = 2\pi\gamma/\tau. \tag{S3.1}$$

where $\gamma$ (kg·m·s$^{-2}$) is the Stokes' drag coefficient of the protein in the solution and $\tau$ (s)is the time constant from a single exponential decay fit to the autocorrelation curve of transmission traces or the first time constant from a double exponential decay fit to the autocorrelation curve.

In free solution, the Stokes' drag coefficient relates to the viscosity of the liquid $\eta$ (Pa·s), and the hydrodynamic radius of the object $r_0$ (m) by the relationship $\gamma_0 = 6\pi\eta r_0$. At the low Reynolds numbers of this system, we can use Faxén's law to calculate the Stokes' drag coefficient considering the distance of the object to the wall $h$ (m) by Equation S3.2:[13,15]

$$\gamma = \frac{6\pi\eta r_0}{\left[1-\frac{9}{16}\left(\frac{r_0}{h}\right)+\frac{1}{8}\left(\frac{r_0}{h}\right)^3-\frac{45}{256}\left(\frac{r_0}{h}\right)^4-\frac{1}{16}\left(\frac{r_0}{h}\right)^5\right]}. \tag{S3.2}$$

We assumed a distance of 1 nm between the trapped object and gold wall for the passivated DNH samples. We then estimated the trapping stiffness of all proteins in this work, as shown in Figure S2. The viscosity of water at 40 °C (the temperature at the hotspot of the DNH at a laser power of 23 mW) is $6.53 \times 10^{-4}$ Pa·s,[16] and the viscosity of an aqueous solution with 10% glycerol (v/v, used in AdK experiments) at 40 °C is $8.75 \times 10^{-4}$ Pa·s.[17] We used a hydrodynamic radius of 2.5 nm for CaM,[18], 3 nm for AdK,[19] and 3.2 nm for Hb,[20] 3.7 nm for Citrate Synthase (CS),[21] 5.5 nm for Immunoglobulin G (IgG),[21] and 5 nm for Glycogen Phosphorylase b (GPb).[22] Figure S2 summarizes the trapping stiffnesses for all proteins we tested in this work, which range in molecular weight from 17 to 180 kDa. We note that with this range of trapping stiffnesses, the DNH in our setup held single proteins in the trap for hours at a flow rate of 0.05 mL min$^{-1}$ (See Figure S9 for details).

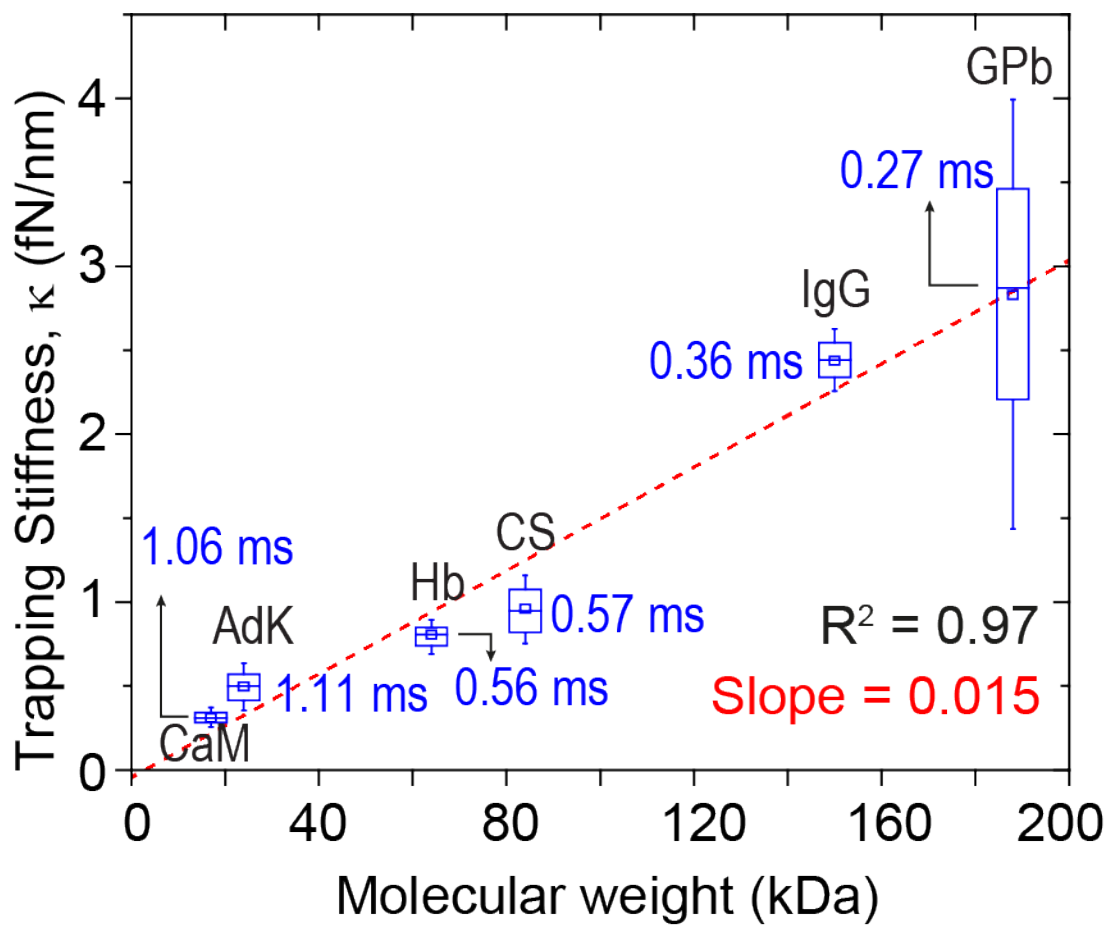

**Figure S2.** Box and whisker plots of trapping stiffness for proteins with different molecular weights, trapped at a laser power of 23 mW. The trapping stiffness was calculated using the time constants from the autocorrelation function of each protein, labeled next to each box (see SI-1 Materials and Methods for details). Median values of trapping stiffness were fit linearly (dashed red line). The boxes span the interquartile range, the middle lines are median values, and the open squares are mean values.

## SI-4 Autocorrelation function of a single hemoglobin in T and R states

We observed a longer decay time in the autocorrelation function of the transmission trace for hemoglobin (Hb) in the R state compared to Hb in the T state (Figure S3). The autocorrelation function of the trace from Hb in the T state fit well to a single-exponential decay with a median time constant, $\tau_1$, of 560 μs. For the transmission signal of Hb in the R state, the autocorrelation function fit well to a double-exponential decay with a fixed first time constant ($\tau_1$) of 560 μs, and a second time constant ($\tau_2$) with a median value of 7.58 ms, that we attribute to conformational fluctuations of Hb in its R-state. See details regarding autocorrelation functions and exponential decay fitting in the SI-1 Materials and Methods.

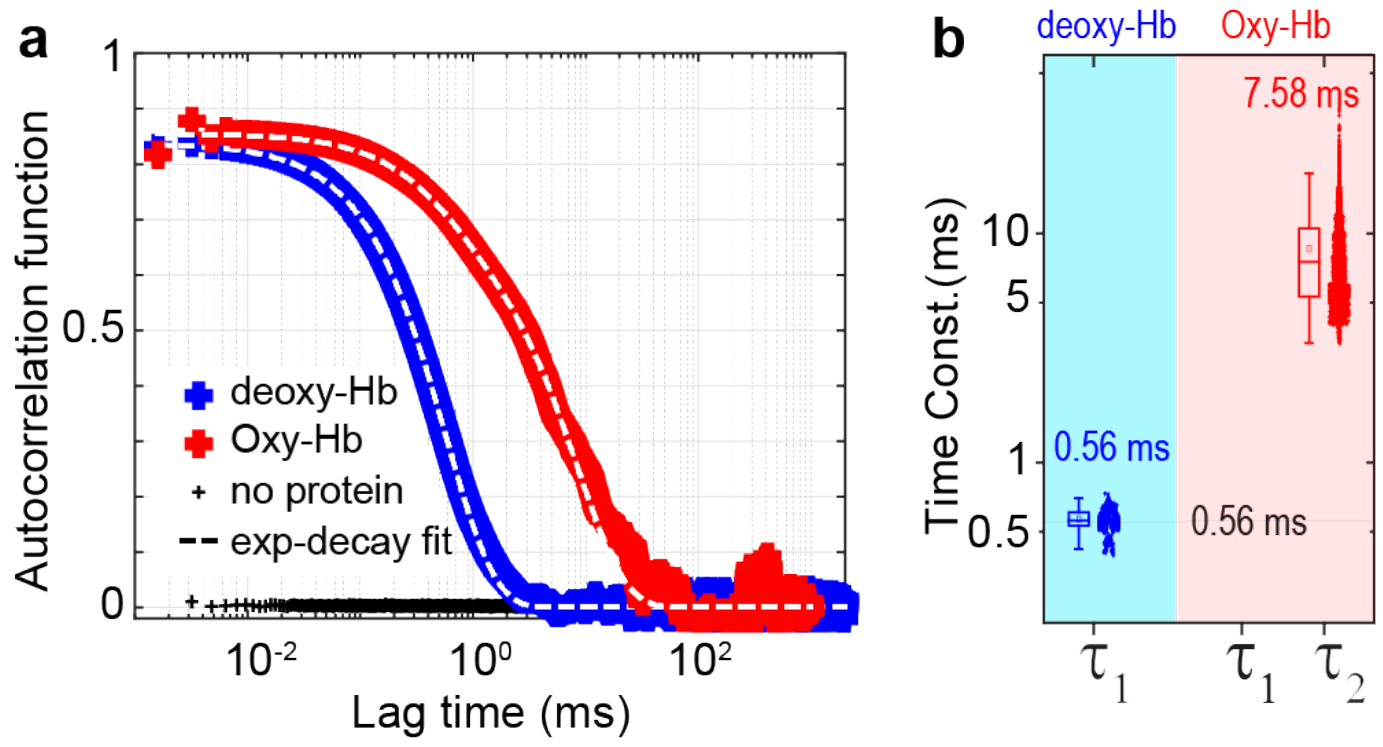

**Figure S3. Autocorrelation function of single-hemoglobin transmission signal shifts in response to oxygen binding. a**, Autocorrelation functions of transmission signals of a double nanohole (DNH) in the absence of protein (black), with Hb trapped in the T (blue) and R (red) states, along with their exponential decay fits (dashed curves). **b**, Time constants from exponential fits to the autocorrelation curves for Hb in the deoxy (T) state and the oxy (R) state, as determined by the exponential decay fits. See materials and methods in SI-1 for details of experiment and fitting of autocorrelation functions.

## SI-5 Transmission through an empty DNH containing oxygen-rich and deoxygenated buffers

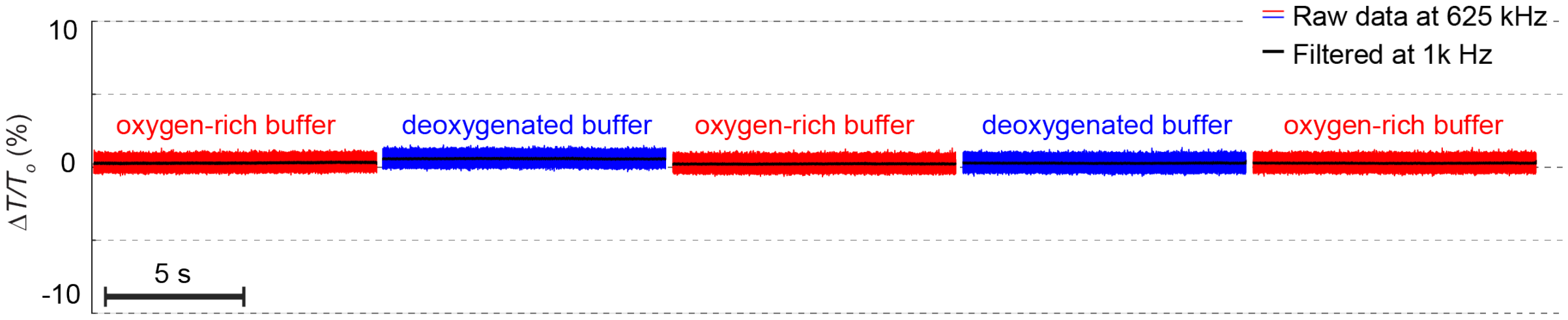

**Figure S4.** Transmission traces of a DNH with an empty hotspot (i.e., no protein trapped) in response to alternating oxygen-rich buffer (red) and deoxygenated buffer (blue) at a laser power of 23 mW (see SI-1 Materials and Methods for details). No changes in transmission signal were observed in response to switching between these two buffers. Data are shown raw (625 kHz) and after digital low-pass filtering with a cutoff frequency of 1 kHz.

## SI-6 Conformational stability of Methemoglobin in response to the addition of oxygen

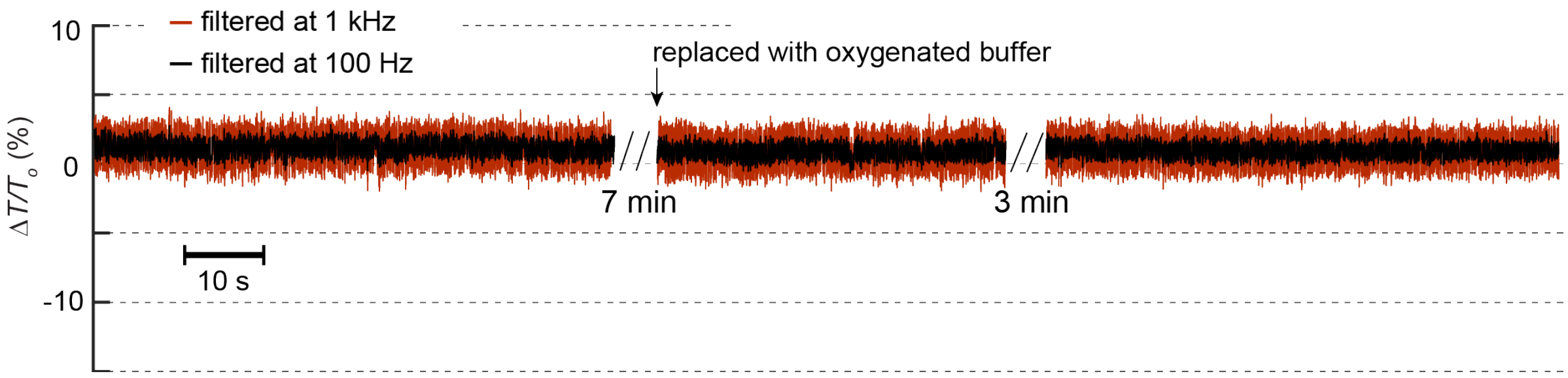

**Figure S5. A single methemoglobin (Met-Hb) in response to the introduction of an oxygenated buffer.** A DNH with a trapped Met-Hb displayed no change in the magnitude or fluctuations of its transmission signal in response to the addition of an oxygenated buffer solution. (see SI-1 Materials and Methods for details). The data were acquired at 625 kHz and are shown after digital low-pass filtering with a cutoff frequency of 1 kHz (dark red) and 100 Hz (black).

## SI-7 Absorbance spectra of Hemoglobin in different solutions

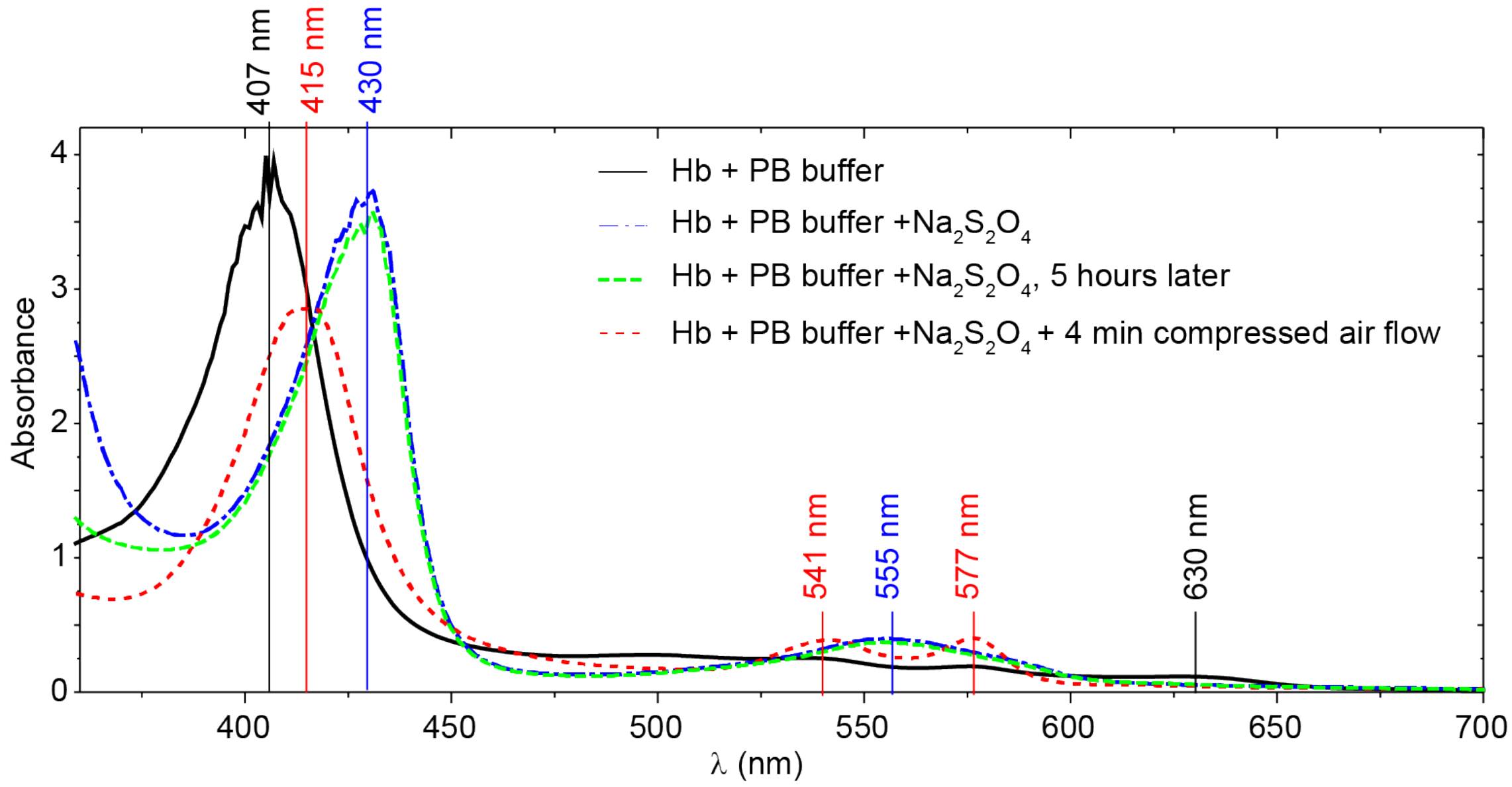

**Figure S6.** UV-Vis absorbance spectra of hemoglobin (Hb) in different conditions. Black curve represents the spectra of 120 μM hemoglobin in phosphate buffer. The spectra indicate that the solution contains primarily methemoglobin (Met-Hb).[54] After mixing the Met-Hb solution with 24 mM sodium hydrosulfite ($Na_2S_2O_4$), the $Fe^{3+}$ in the heme group of methemoglobin was reduced to $Fe^{2+}$ such that the protein was in the deoxygenated state of hemoglobin (deoxy-Hb), as represented by the blue curve. The single peak at 555 nm in the absorbance spectra indicates that the solution contains deoxy-Hb. The solution contained deoxy-Hb even after 5 hours of storage at room temperature (green curve). The red curve represents the spectra of a freshly mixed deoxy-Hb solution after blowing compressed air into the solution for 4 min. Two absorbance peaks at 541 nm and 577 nm indicate that the hemoglobin in the solution is oxygen-loaded hemoglobin (oxy-Hb).[54]

## SI-8 Temperature in trapping site

To evaluate the temperature in the plasmonic hotspot of the DNH as a function of laser intensity, we measured the change in ionic current though a nanopore in the gap of the DNH structure (Figure S7). To do so, we prepared a fused silica chip with a 30-nm thick freestanding $SiN_x$ membrane,[23] and then deposited a 5 nm Ti film and 100 nm gold film on top of the $SiN_x$. We milled a nanopore with a diameter of 32 nm through the $SiN_x$ membrane in the gap of the double nanohole structure (Figure S7a inset). When an electrolyte solution fills a nanopore, the ionic current through the nanopore is determined by Equation S8.1:[24]

$$I = \frac{V}{\rho}\left(\frac{l_p}{\pi R_p{}^2} + \frac{1}{2R_p}\right)^{-1} \quad \text{(S8.1)}$$

Here the applied voltage across the nanopore, $V$ (V), is 0.2 V, the nanopore's length, $l_p$ (m), is 15 nm (see SI-2 for details), the radius of the nanopore, $R_p$ (m), is 16 nm, and the resistivity of the electrolyte solution, $\rho$ (Ω/m), is 0.046 Ω/m for the 2 M KCl aqueous electrolyte solution that we used at 21 °C. Figure S7a shows that the measured ionic current across the membrane reversibly changed by increasing and decreasing the power of a laser focused on the DNH-nanopore structure. This change in resistivity originates from the temperature-dependent reduction of electrolyte viscosity by the relationship $\rho = C\eta(T)$,[24–27] where the viscosity $\eta$ (Pa·s) can be calculated according to Equation S8.2:[28]

$$\eta = (2.414 \times 10^{-5}\ \mathrm{Pa \cdot s}) \times 10^{\left(\frac{247.8\mathrm{K}}{T - 140\mathrm{K}}\right)}. \quad \text{(S8.2)}$$

Figure S7b shows the temperature of the DNH in response to laser illumination. We determined the temperature in response to laser-induced heating based on the current through the nanopore (Figure S7b), and simulated the temperature using finite-element simulation in COMSOL Multiphysics (Figure S7c).[24,29] As reported by Jiang *et al.*, the laser-induced heating of nanostructures with nanohole diameters below 350 nm originates from photon absorption by the gold layer and subsequent energy dissipation.[29] Based on this assumption, we performed a COMSOL Multiphysics simulation using the Transient Heat Transfer model as described in previous work.[24] We considered the main heat source to be the absorbance of the laser light by both the electrolyte solution and the gold film. Figure S7c shows the temperature profile of a Gaussian-shaped laser beam with a diameter of 1.44 µm and laser power of 23 mW incident to the DNH-nanopore. The absorption coefficient is 4.4372 $m^{-1}$ for water at 852 nm,[30] and $7.87 \times 10^7$ $m^{-1}$ for gold at 852 nm.[31] The peak temperature of 40 °C predicted by the simulation for a DNH excited by a 23 mW laser matches the trend of temperatures observed experimentally using the ionic current through the nanopore. We chose a laser power of 23 mW because it made it possible to trap proteins with minimal unintended escapes from the trap while the laser was on and because it led to a temperature in the DNH that was close to 37 °C, the body temperature of humans.

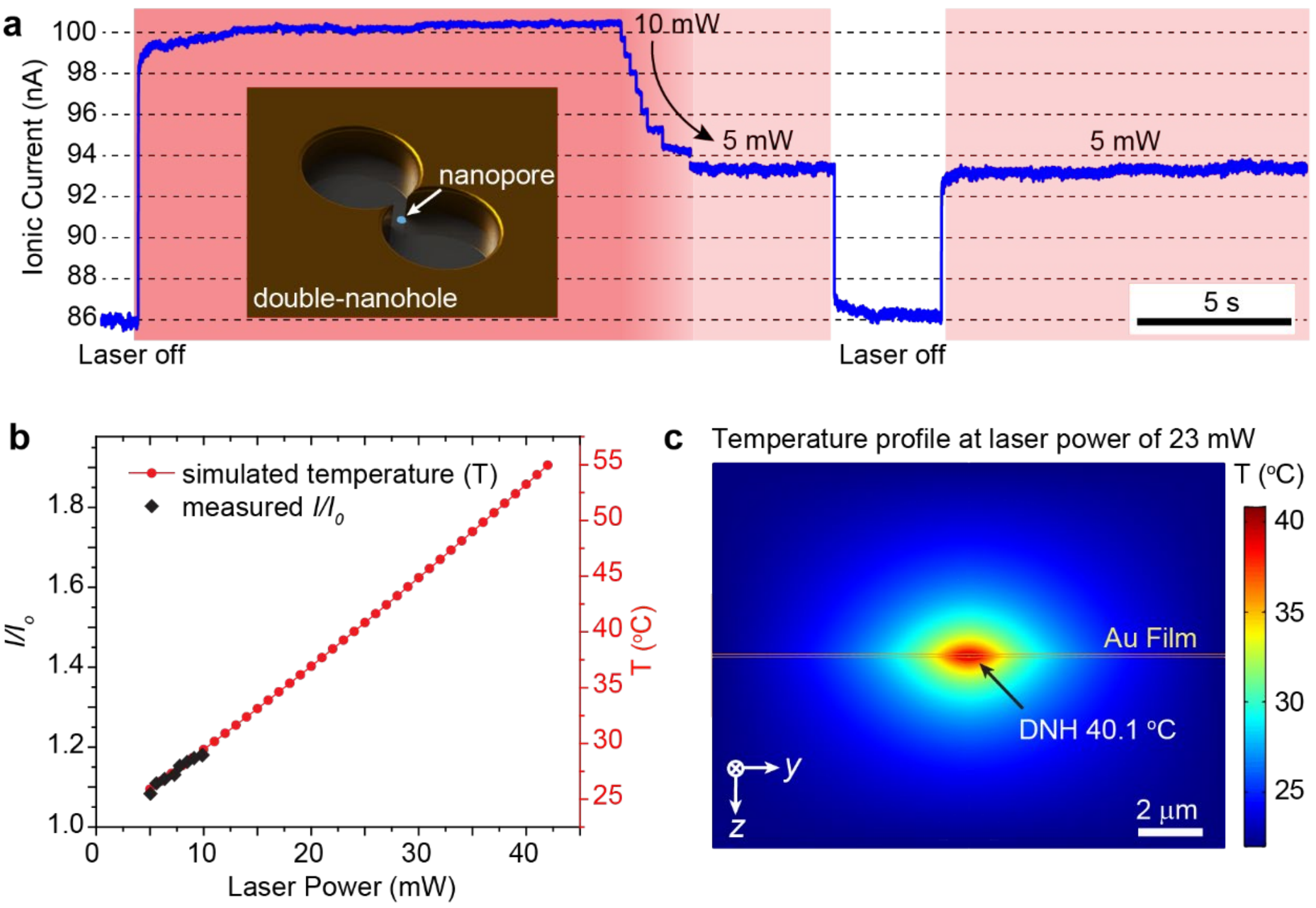

**Figure S7. Laser-induced heating in the DNH sample**. **a**, At an applied voltage of 0.2 V, the ionic current measured through a nanopore across the $SiN_x$ membrane decreased with decreasing laser power and returned to a value within ±2% of its baseline value within 1 s in the absence of laser illumination, and 90% of this change happened within 70 µs. Ionic current traces were acquired at a sampling rate of 20 kHz by using an amplifier (e4, Elements srl., Italy). **b**, Comparison of experimental data (black squares) with finite element simulation (red circles) of ionic current as a function of laser power, represented as the ratio between the current values in the presence ($I$) and absence ($I_0$) of laser illumination. The right *y*-axis shows the temperature in the nanopore as estimated from the conductance increase in response to increasing laser power. **c**, Finite element simulation of the temperature profile within the DNH-nanopore structure when illuminated at a power of 23 mW.

## SI-9 Stable conformational state of apo-CaM at 47 ˚C

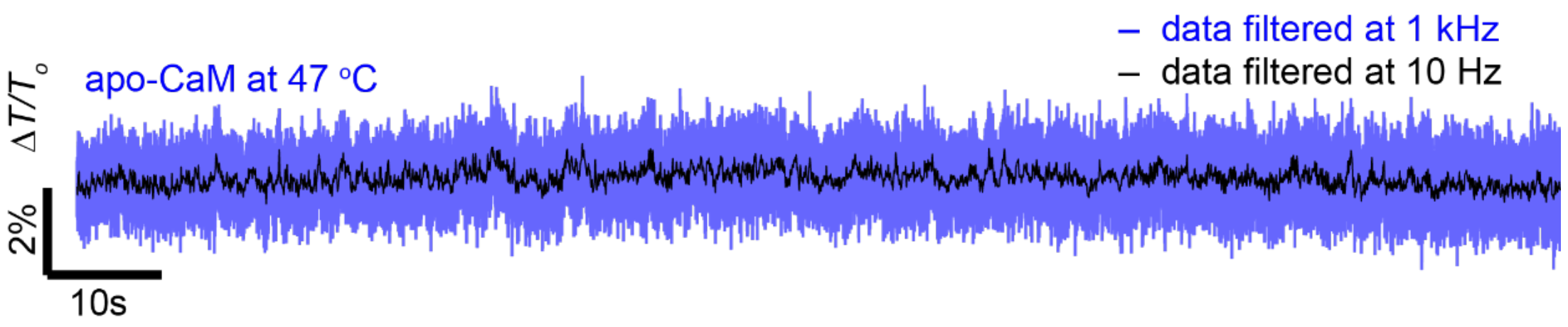

**Figure S8.** Transmission versus time trace of a single *apo*-CaM trapped in the DNH at 47 °C for a duration of 2 min.

## SI-10 Example of continuous trapping of a single calmodulin protein for more than 2 hours

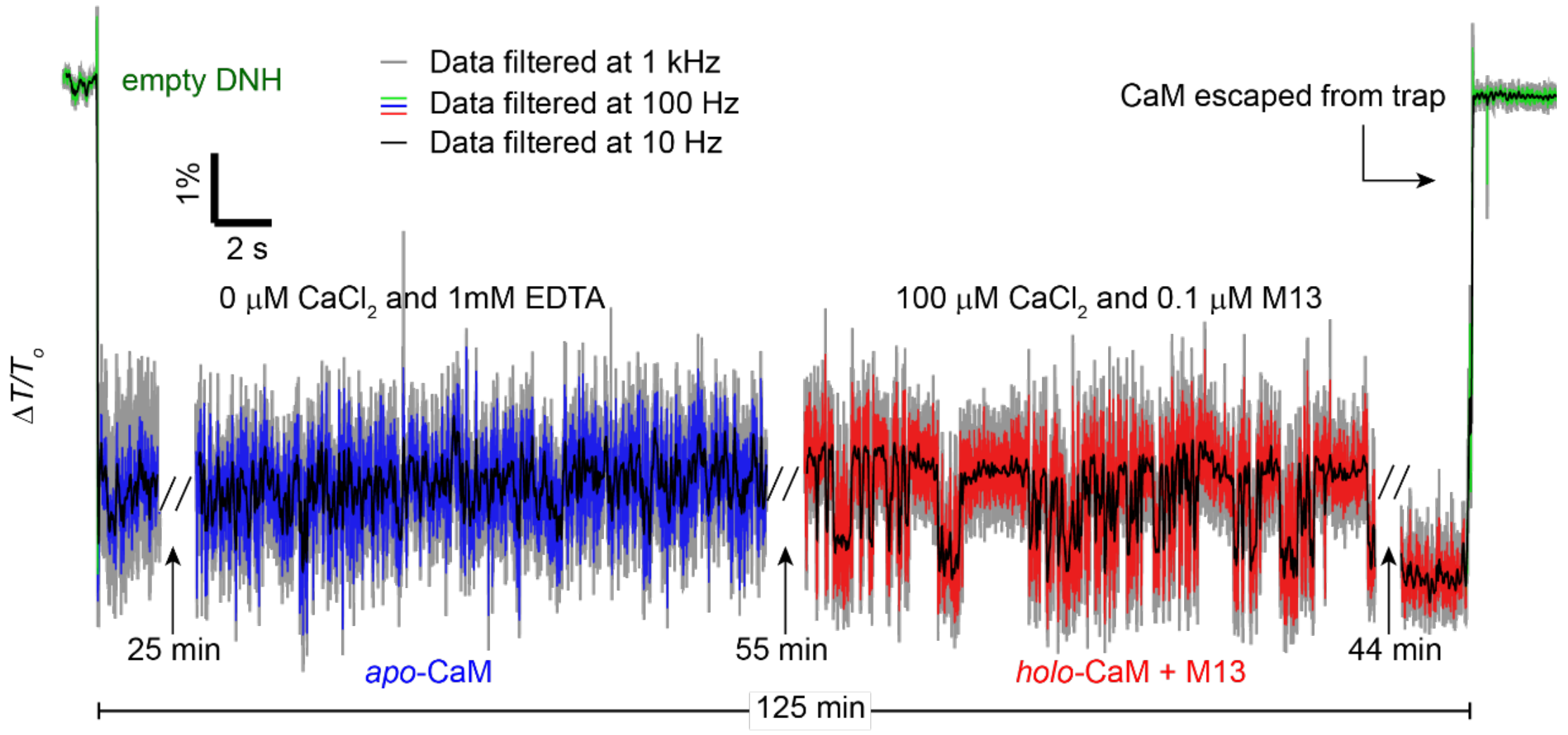

**Figure S9** Continuous transmission trace of a single calmodulin (CaM) protein trapped in the hotspot of a DNH at a laser power of 23 mW (40 ℃) for more than two hours. During this time, we exposed the trapped protein to calcium-free buffer (blue) and then exchanged this buffer with calcium rich buffer containing 100 nM M13 peptide (red) (Ac-KRRWKKNFIAVSAANRFKKISSSGAL-NH2, synthesized by Bio Synthesis company, Bio-Synthesis, Inc.). The steep transitions in the transmission signal with an amplitude of approximately 2% after the addition of M13 indicate binding and unbinding of M13 to CaM, as expected.[32,33] Data were acquired at 625 kHz and are shown after digital low-pass filtering with a cutoff frequency of 1 kHz (grey traces), 100 Hz (color coded traces) and 10 Hz (black traces). The data show that CaM escaped the trap after approximately 125 min. (see SI-1 Materials and Methods for details).

## SI-11 Kinetic model for adenylate kinase

To provide further evidence for the hypothesis that the observed signal fluctuations are due to the conformational changes of the trapped protein, we performed a theoretical analysis of the reaction kinetics of AdK in the presence of increasing ADP substrate concentrations. We considered the following aspects:

i) The trapped protein is not diffusing and the ligands (substrate and product) are under flow. This flow biases the diffusion of substrate and products.
ii) The substrate is continuously supplied. Therefore, concentration of substrate is approximately constant.
iii) The products are continuously removed. Therefore, their concentrations are vanishing.
iv) Ligand (substrate or product) diffusion will occur mostly from above the DNH sample. As the protein is trapped inside the DNH, this location reduces the protein-ligand collision frequency in comparison to the freely diffusing case.

AdK undergoes conformational changes upon binding substrates (ADP) and releasing products (ATP, AMP) during its catalytic activity. In the absence of substrates, AdK is in its open ($o$) conformation, with one bound substrate the protein is in its partially closed ($pc$) conformation, and upon binding two substrates it switches to its fully closed (*fc*) conformation.[2, 5]

The reaction of AdK with ADP (continuously supplied) to produce AMP and ATP (continuously removed) is modelled as shown in Figure S10a. In the first two steps, AdK binds two ADP molecules in sequence, converts them into one ATP and one AMP molecule, and subsequently releases those products. Figure 3b shows that we never observed a single step change from $fc$ to $o$, indicating that the products AMP and ATP are never released simultaneously but rather sequentially. This observation agrees with previous work that ATP release is the rate-limiting step in this release process, and that the ATP lid and AMP lid of AdK open and close independently.[34] Under these assumptions, the expected reaction mechanism simplifies as shown in Figure S10a. Four types of events can be identified, each of which can be characterized by a kinetic constant.

The constants $k_1$ and $k_2$ (Figure S10a) are linked to the frequencies at which the ADP molecules bind AdK. From collision theory we know that the number of successful (leading to a reaction) collisions per unit time ($\nu_c$) between two reacting species is[36]

$$\nu_c = N_i N_j \frac{K}{V} \mathrm{e}^{-\frac{\Delta E_a}{k_\mathrm{B} T}}, \tag{S11.1}$$

where $N$ is the number of particles of each colliding species, $V$ ($\mathrm{m}^3$)is the solution volume, $K$($\mathrm{m}^3 \cdot \mathrm{s}^{-1}$) is a constant that accounts for the size, diffusivity and relative orientation of the molecules, $\Delta E_a$ (J) is the activation energy of the reaction, $k_\mathrm{B}$ (J/K) is Boltzmann's constant and $T$ $(K)$ is temperature.

For a single molecule in the gap of a DNH, the frequency of successful collisions can be interpreted as the inverse of the average lifetime $\bar{t}_n$(s) of a certain state $n$. Considering that a single molecule is trapped ($N_i = 1$) and substituting $N_j/V = C_j$ yields[a]

$$\nu_c = \frac{1}{\bar{t}_n} = C_j K \mathrm{e}^{-\frac{\Delta E_a}{k_\mathrm{B} T}}. \tag{S11.2}$$

Expressing the concentration of $j$ as a molar concentration $[j]$ by means of the Avogadro constant $N_\mathrm{av}(\mathrm{mol}^{-1})$, Equation S11.2 becomes

$$\nu_c = \frac{1}{\bar{t}_n} = [j] N_\mathrm{av} K \mathrm{e}^{-\frac{\Delta E_a}{k_\mathrm{B} T}} = [j] k_i \tag{S11.3}$$

Equation S11.3 applies to both the binding of the first and the second ADP molecules to AdK:

$$\bar{t}_1 = \frac{1}{k_1[\mathrm{ADP}]}, \bar{t}_2 = \frac{1}{k_2[\mathrm{ADP}]} \tag{S11.4}$$

Equation S11.4 implies that the average times for AdK to undergo the transition from open ($o$) to partially closed ($pc$) and from $pc$ to fully closed (*fc*) are expected to scale linearly with the inverse of ADP concentration with slope $1/k_1$ and $1/k_2$ respectively. Conversely, all other processes described in Figure S10a are unimolecular dissociations that are not expected to involve molecular collisions, and are therefore not expected to depend on ADP concentrations:

$$t_{-1} = \frac{1}{v_{-1}} = \frac{1}{k_{-1}},\ t_{-2} = \frac{1}{v_{-2}} = \frac{1}{k_{-2}},\ t_3 = \frac{1}{v_3} = \frac{1}{k_3},\ t_4 = \frac{1}{v_4} = \frac{1}{k_4}. \quad \text{(S11.5)}$$

[a] Under the experimental conditions described in this work, the substrate constantly flows in the direction parallel to the gold surface. Therefore, Equation S11.2 requires a further correction to allow quantitative prediction of $\bar{t}_n$.[37] This step, however is not necessary and out of scope of the present discussion.

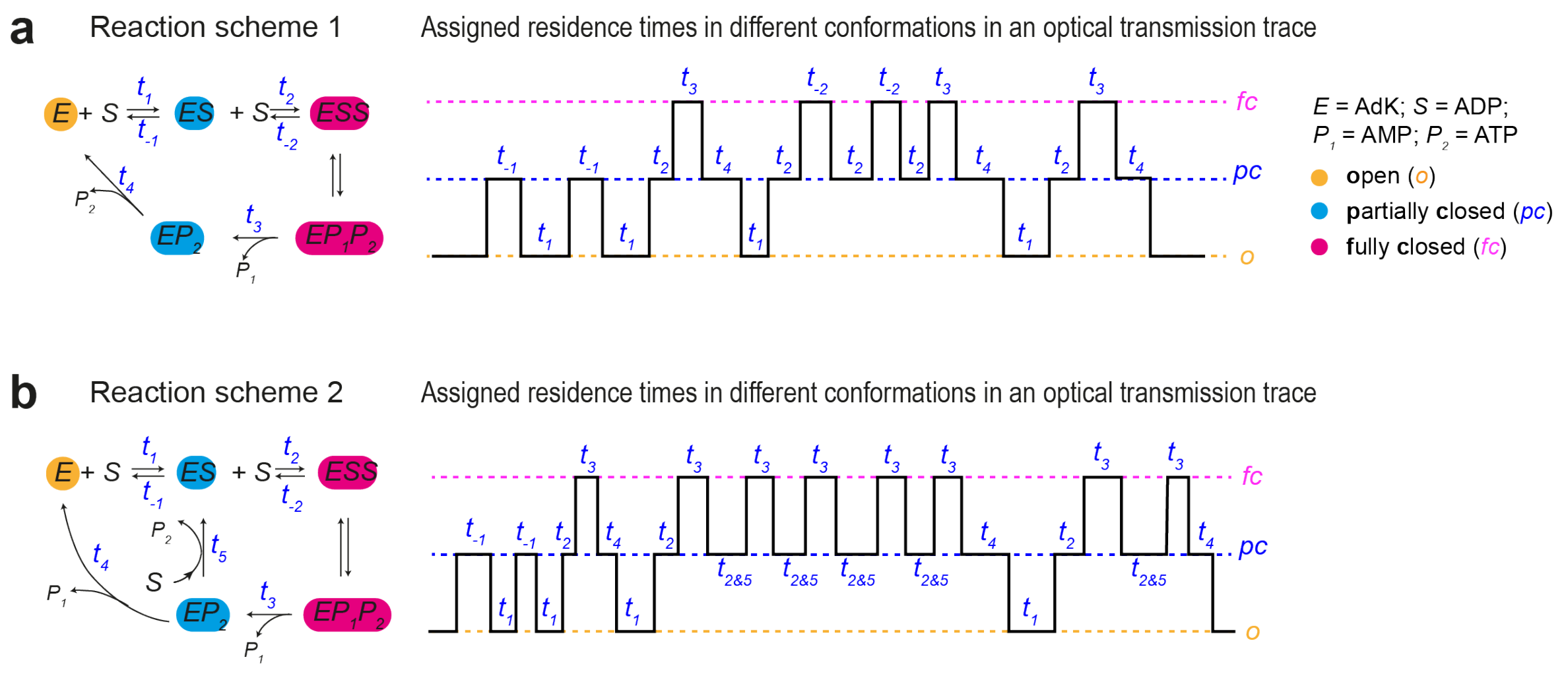

**Figure S10. Schematic representation of the conformational kinetics of AdK with ADP and the assigned residence times in different conformations in an optical transmission trace. a**, Reaction scheme 1, in which the enzyme is required to sample its open conformation to carry out the catalytic reaction. **b**, Reaction scheme 2, in which the enzyme does not need to sample its open confirmation to complete a full turnover cycle; instead sampling of the partially closed conformation can be sufficient to carry out a catalytic cycle. The black curves represent idealized transmission traces of a single trapped AdK transitioning between its various conformational states according to these two reaction schemes. The three dashed lines represent transmission levels corresponding to the three conformations of AdK with the open and most elongated conformation represented by the lowest transmission level and the closed and most compact conformation represented by the highest transmission level.

To compare these predictions with the experimental data, we developed an algorithm to identify the order in which AdK samples its various conformations, and its residence time in each conformation. This algorithm first performed a step fitting of the optical transmission traces. The fitting identifies three different intensity levels corresponding to three expected AdK conformations, as shown in Figure 3d in the main text. To identify conformational dynamics, we took advantage of additional information provided by the single molecule data. For instance, for the *pc* conformation, if the previous conformation is *fo* and the next conformation is *fc*, we identify the lifetime of this level as $t_2$, corresponding to the binding of a second ADP. If, on the other hand, both the previous and the next conformations are *fo*, we assign this level as $t_{-1}$, corresponding to the dissociation of the first ADP. We extracted two arrays from the step-fitted trace, one defining the observed sequence of conformations (e.g. *c* = [o, pc, fc, pc, fc, …])), and one reporting the residence time of each conformation (e.g. $t$ = [3, 5, 9, 10, 5,…] ms). Table 1 lists the logic that the algorithm uses to classify a certain $t_l$ based on $c_{l-1}, c_l, c_{l+1}$ and, if needed, $c_{l+2}$. Based on this logic, we assigned each step in the step-fitted transmission trace to different binding times, as illustrated in Figure S10. Once we classified all $t_l$ into the $n$ classes, we fit the distributions of $t_n$ within each class with the non-linear least squares regression method to a single exponential decay function to determine the average $\bar{t}_n$value.

**Table S1**. Sequences of $c_i$ used to classify the various $t_i$ into $n$ categories.

| $t_n$ | $c_{i-1}$ | $c_i$ | $c_{i+1}$ | $c_{i+2}$ |
|---|---|---|---|---|
| $t_1$ | *pc, fc* | *fo* | *pc* | - |
| $t_{-1}$ | *fo* | *pc* | *fo* | - |
| $t_2$ *or* $t_{2\&5}$ | *fo* | *pc* | *fc* | - |
| $t_{-2}$ | *pc* | *fc* | *pc* | *fc* |
| $t_3$ | *pc* | *fc* | *pc* | *fo* |
| $t_4$ | *fc* | *pc* | *fo* | - |

While assumptions in previous reports[34,38] inspired the reaction scheme presented in Figure S10a, we have noted that at high substrate concentrations, *e.g.* 1 mM ADP (Figure 3b in the main text), the enzyme alternates for extended periods between the transmission levels of *fc* and *pc*, while passing through the open level *o* too rarely to rationalize the reported high turnover rates at this high substrate concentration.[34,35] This observation, together with the dependence of unbinding of ATP on the concentration of ADP (see main text) led us to the reaction scheme in Figure S10b, which supposes that before releasing the second product ATP, the AMP lid of AdK binds ADP (a reaction that occurs faster at increasing ADP concentrations), which in turn accelerates the release of ATP (Figure 3f, main text). Under this assumption, the time constant $t_2$ represents the total time of the enzyme releasing an ATP and binding the second ADP, hence written as $t_{2\&5}$ according to the reaction scheme in Figure S10b. Considering $t_{2\&5}$ and $t_3$ to be a full cycle leads to an estimated turnover rate of 200 $s^{-1}$, in agreement with AdK's reported turnover rate of 116 to 210 $s^{-1}$.[34,39]

## SI-12 Level occupancies and probabilities of various conformational states of subunits of citrate synthase

**Table S2.** Level occupancies from 5-peak Gaussian fit to 100 kHz data and conformation probabilities for subunits of citrate synthase in the absence of OAA and in the presence of 150 µM OAA and various AcCoA concentrations.

| OAA (µM) | AcCoA (µM) | Assigned levels | | | | | Probability of states | | |
|---|---|---|---|---|---|---|---|---|---|
| | | *L1 (c-c)* | *L2 (c-i)* | *L3 (i-i, c-o)* | *L4 (o-i)* | *L5 (o-o)* | $P_o$ | $P_c$ | $P_i$ |
| 0 | 0 | 0 | 0 | 0 | 0.149 | 0.840 | 0.92 | 0 | 0.08 |
| 150 | 0 | 1.0 | 0 | 0 | 0 | 0 | 0 | 1 | 0 |
| 150 | 4 | 0.176 | 0.649 | 0.188 | 0 | 0 | 0 | 0.59 | 0.43 |
| 150 | 6 | 0.092 | 0.465 | 0.429 | 0.016 | 0 | 0.005 | 0.33 | 0.66 |
| 150 | 10 | 0.034 | 0.227 | 0.393 | 0.273 | 0.076 | 0.27 | 0.20 | 0.54 |
| 150 | 36 | 0.029 | 0.229 | 0.389 | 0.252 | 0.102 | 0.28 | 0.20 | 0.53 |
| 150 | 400 | 0.016 | 0.116 | 0.284 | 0.384 | 0.190 | 0.45 | 0.14 | 0.42 |
| 150 | 1000 | 0.056 | 0.194 | 0.332 | 0.276 | 0.138 | 0.39 | 0.24 | 0.37 |

## SI-13 Kinetic model for Citrate Synthase

This section will show step-by-step that the sequence of conformational changes and the relative occurrence (probability) of each conformation obtained by the plasmonic optical trapping analysis are related to the rate of the Citrate Synthase (CS) enzymatic reaction, i.e. the conversion of OAA and AcCoA into citrate and CoA.

### SI-13.1 Conformation probabilities from the transmission signal

We define the probability of a certain protein conformation, $P_j$, as the amount of time spent in that conformation, $t_j$(s), divided by the total sampling time, $t_{tot}$ (s),

$$P_j = \frac{t_j}{t_{tot}}. \qquad \text{(S13.1)}$$

It is generally assumed, that CS has two independent active subunits, each of which can adopt two conformations, defined as *open* (*o*) and *closed* (*c*). This structure leads to 3 possible combinations, which are *o-o*, *o-c*, and *c-c*. From the transmission signals, however, we observed that CS samples at least 5 levels while it is active, as illustrated in Figure 4d in the main text. Table S3 lists the probabilities occupying of each level determined by step fit of the transmission traces to 5 levels. The observation of more than three levels in the transmission signal suggests the presence of an additional *intermediate* (*i*) conformation for each subunit. With three conformations per subunit, however, we expect 6 different combinations: *o-o*, *o-i*, *i-i*, *o-c*, *i-c*, and *c-c*. The observation of only 5 levels indicates that two of these conformations, *i-i* and *o-c* (see main text), give a similar response in the transmission signal and therefore cannot be distinguished. To overcome this problem, we used an iterative procedure described in section SI-13.2.

**Table S3.** Normalized level occupancies and refined conformation probabilities (obtained as described in section SI-13.2) for citrate synthase in the presence of 150 μM OAA and various AcCoA concentrations.

| AcCoA | Assigned levels | | | | | Refined | | |
|---|---|---|---|---|---|---|---|---|
| (μM) | *L1 (c-c)* | *L2(c-i)* | *L3 (i-i)* | *L4 (o-i)* | *L5 (o-o)* | $P_o$ | $P_c$ | $P_i$ |
| 0 | 1 | 0 | 0 | 0 | 0 | 0 | 1 | 0 |
| 4 | 0.3016 | 0.657 | 0.0412 | 0 | 0 | 0.0000 | 0.6302 | 0.3697 |
| 6 | 0.1721 | 0.681 | 0.145 | 0.0003 | 0 | 0.0003 | 0.5133 | 0.4863 |
| 10 | 0.0525 | 0.401 | 0.462 | 0.1298 | 0.0066 | 0.1461 | 0.3187 | 0.5350 |
| 36 | 0.0003 | 0.121 | 0.631 | 0.2339 | 0.0139 | 0.2739 | 0.2037 | 0.5222 |
| 400 | 0.0330 | 0.2233 | 0.4716 | 0.2876 | 0.0174 | 0.2833 | 0.2674 | 0.4492 |
| 1000 | 0.0834 | 0.2927 | 0.3483 | 0.2401 | 0.0356 | 0.2609 | 0.3351 | 0.4038 |

### SI-13.2 Iterative procedure to refine the conformation probabilities

By assigning each transmission level to a certain combination of conformations, the time that a given subunit spends in each of its three conformations can be written as Equations S13.2-S13.4:

$$t_o = t_{o-o} + \frac{1}{2}t_{o-i} + \frac{1}{2}t_{o-c}, \qquad \text{(S13.2)}$$

$$t_c = t_{c-c} + \frac{1}{2} t_{c-i} + \frac{1}{2} t_{o-c}, \quad \text{(S13.3)}$$

$$t_i = t_{i-i} + \frac{1}{2} t_{o-i} + \frac{1}{2} t_{i-c}, \quad \text{(S13.4)}$$

Using Equation S13.1, we can calculate the probability of each conformation as

$$P_o = \frac{1}{t_{\text{tot}}} \left( t_{\text{o-o}} + \frac{1}{2} t_{\text{o-i}} + \frac{1}{2} t_{\text{o-c}} \right), \quad \text{(S13.5)}$$

$$P_c = \frac{1}{t_{\text{tot}}} \left( t_{\text{c-c}} + \frac{1}{2} t_{\text{c-i}} + \frac{1}{2} t_{\text{o-c}} \right), \quad \text{(S13.6)}$$

$$P_i = \frac{1}{t_{\text{tot}}} \left( t_{i-i} + \frac{1}{2} t_{o-i} + \frac{1}{2} t_{c-i} \right). \quad \text{(S13.7)}$$

The transmission signal in our experiment provides only the sum of $t_{i-i}$ and $t_{o-c}$ as $t_{L3}$ rather than their individual values. Hence, we replace $t_{i-i}$ with $t_{L3} - t_{\text{o-c}}$, which leads to Equations S13.8 - S13.10

$$P_{\text{o}} = P_{\text{L1}} + \frac{1}{2} P_{\text{L2}} + \frac{1}{2} P_{\text{o-c}}, \quad \text{(S13.8)}$$

$$P_{\text{c}} = P_{\text{L5}} + \frac{1}{2} P_{\text{L4}} + \frac{1}{2} P_{\text{o-c}}, \quad \text{(S13.9)}$$

$$P_i = P_{\text{L3}} + \frac{1}{2} P_{\text{L2}} + \frac{1}{2} P_{\text{L4}} - P_{\text{o-c}}. \quad \text{(S13.10)}$$

By assigning an arbitrary initial value to $P_{o-c}$, we can calculate the initial probabilities of $P_o$, $P_i$ and $P_c$. Since the activity of one subunit of CS is independent of the activity of the other subunit, the probability to sample the *o-c* conformation is

$$P_{o-c} = 2 P_o P_c, \quad \text{(S13.11)}$$

The iterative procedure to refine the probability values is as follows:

1. The initial values for $P_{\text{o}}$, $P_{\text{c}}$ and $P_{\text{I}}$ are calculated from (S13.8)-(S13.10) using an arbitrary $P_{o-c}$ initial value. We used $P_{o-c} = 0.1$.
2. $P_{o-c}$ is calculated from equation (S13.11).
3. $P_o$, $P_c$ and $P_i$ are re-calculated from (S13.8)-(S13.10) using the new $P_{o-c}$ value.
4. Steps 2 and 3 are repeated until the difference between two consecutive iterations of $P_{\text{o-c}}$ is smaller than $10^{-9}$.

We used this procedure to calculate the refined conformation probabilities reported in Table S3. Knowing the probability of the protein in a certain conformation is important for understanding the activity of that protein, as shown in the following sections.

### SI-13.3 Deriving the reaction scheme from transmission signals

We analysed the level occupancy and developed an algorithm that analyses the sequence between the five levels present in the step-fitted CS transmission signal in order to reveal which transitions are allowed and which ones are not. In the absence of any substrate, CS transitions between *L3*, *L4* and *L5*, indicating that each of the subunits is transitioning between its open and intermediate conformations (Fig. S11a). In the presence of 150 µM OAA, on the other hand, the protein only populates *L1*, corresponding to the *c-c* configuration (see Fig. S11b and Table S3).

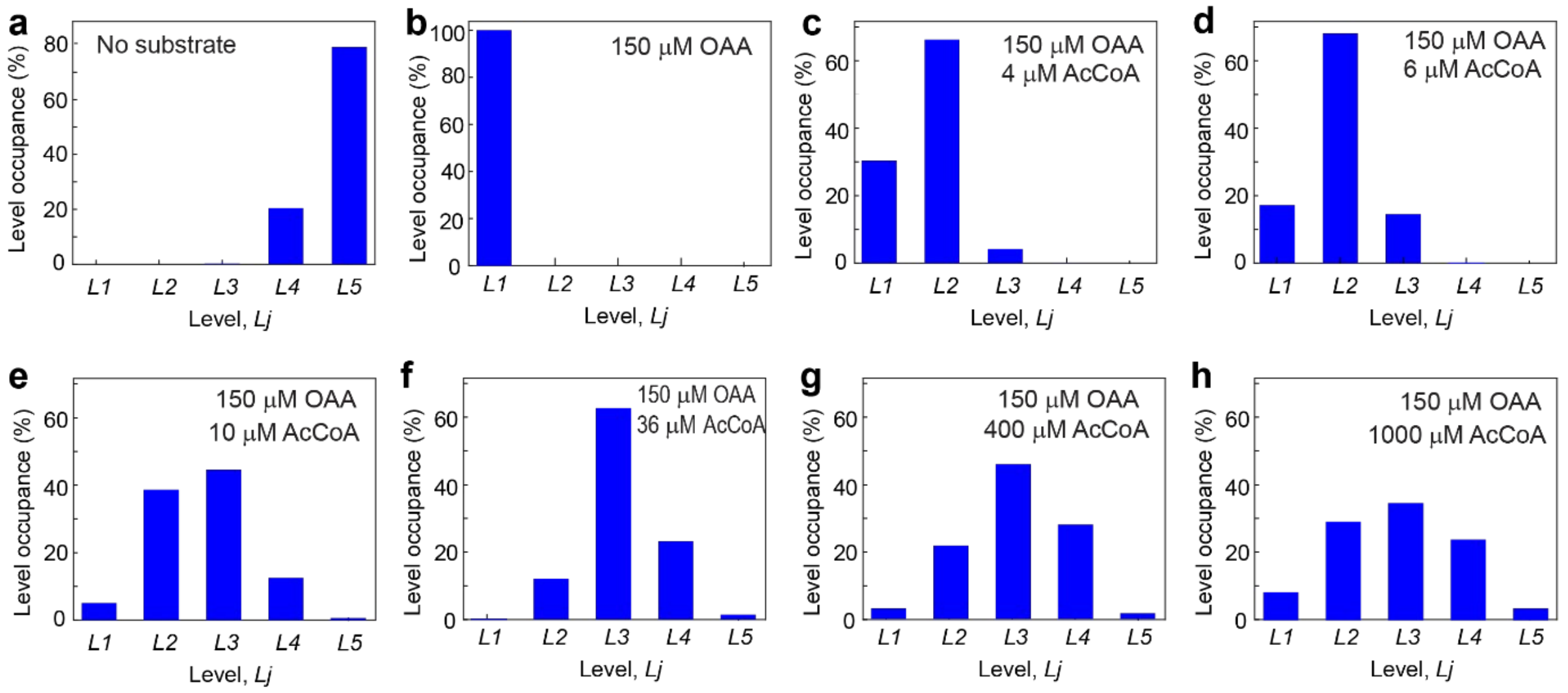

**Figure S11.** Level occupancies of CS, in the absence of any substrate **a**, in the presence of only 150 µM OAA **b,** and with 4, 6, 10, 36, 400, and 1000 µM AcCoA (**c-g**). The levels were obtained from the 5-step fitting of transmission signals of trapped citrate synthase (see Figure 4d in the main text and Materials and Methods in SI-1 for details).

In the presence of low AcCoA concentrations, such as 4 µM and 6 µM (Figure S12 and Table S3), the protein starts to cycle between *L1*, *L2* and *L3*, meaning that the catalytic cycle of CS only involves the *i* and *c* configurations and is not required to transition through the *o* conformation. Figure S12 shows that only the transitions between $L_1$ and $L_2$ as well as between $L_2$ and $L_3$ are occurring in these two conditions.

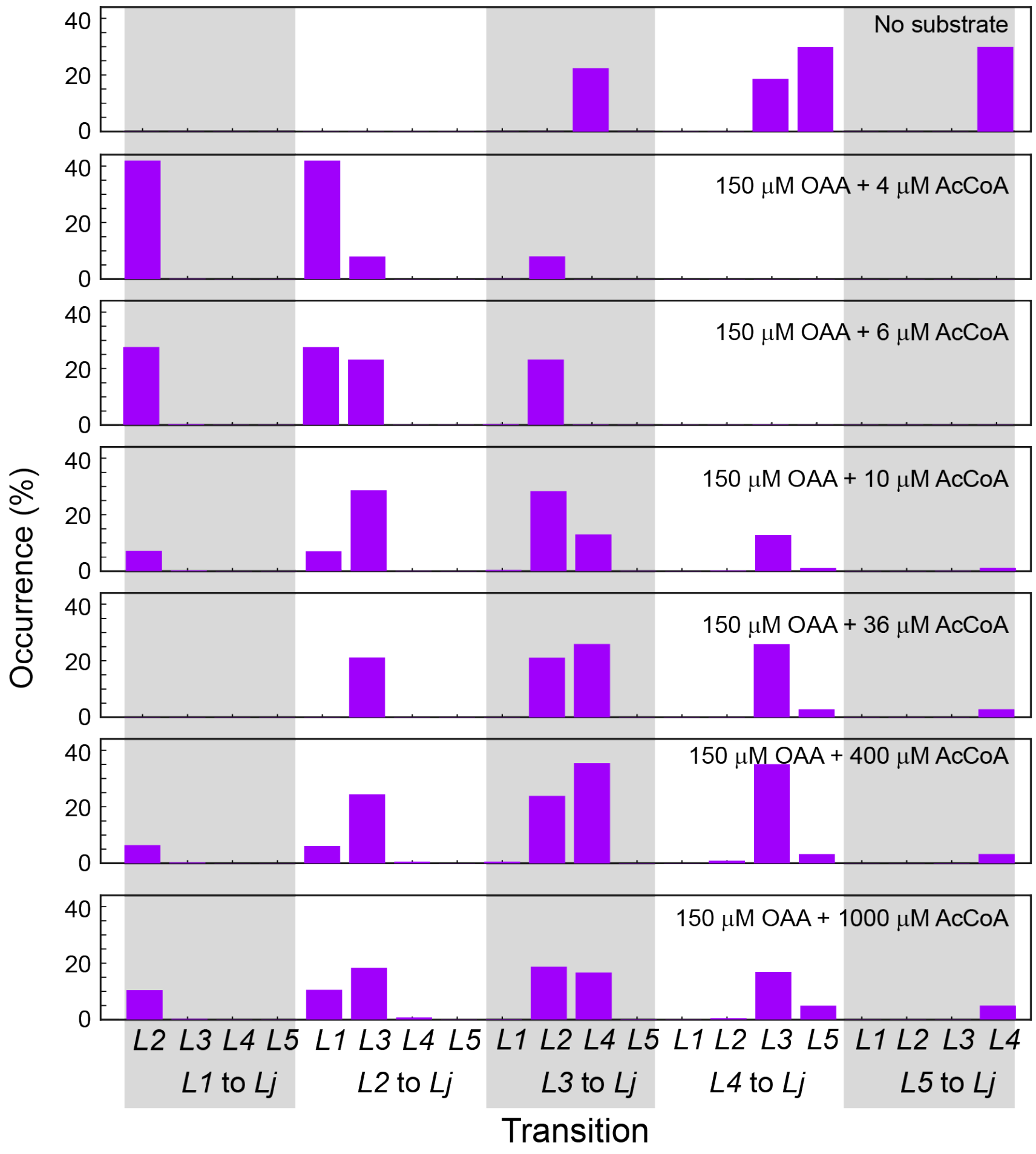

**Figure S12.** Relative occurrence of specific level-to-level transitions for citrate synthase in the absence of any substrate, and in the presence of 150 μM OAA with 4, 6, 10, 36, 400, and 1000 μM AcCoA.

At relatively high AcCoA concentrations (10, 36, 400, 1000 μM, Figure S11e-g and Table S3), the protein mostly cycles between *L1*, *L2* and *L3*. In addition, *L1* reaches a minimum at an AcCoA concentration of 36 μM and then progressively increases. *L5* also becomes more prominent with increasing AcCoA concentration but it approaches a plateau at the highest AcCoA concentrations. This observation suggests that the protein favours its *c* conformation in the presence of AcCoA.[40] Citrate synthase is substrate inhibited by AcCoA.[41] The observation that the enzyme increasingly populates the *c* conformation at the expense of the *i* conformation at high concentrations of AcCoA suggests that inhibition occurs when AcCoA binds to the intermediate *i* conformation of the enzyme before OAA, bringing it into a closed inactive (ci) form not directly distinguishable from the closed active (ca) form. Since we only observed transitions between *L1* and *L2*, *L2* and *L3*, *L3* and *L4* as well as *L4* and *L5* (Fig. S12), the protein explores the levels in sequence, without jumping conformations. Thus, each subunit needs to pass through the *i* conformation when transitioning between *c* and *o* conformations. All these observations suggest the reaction scheme depicted in Figure S13.

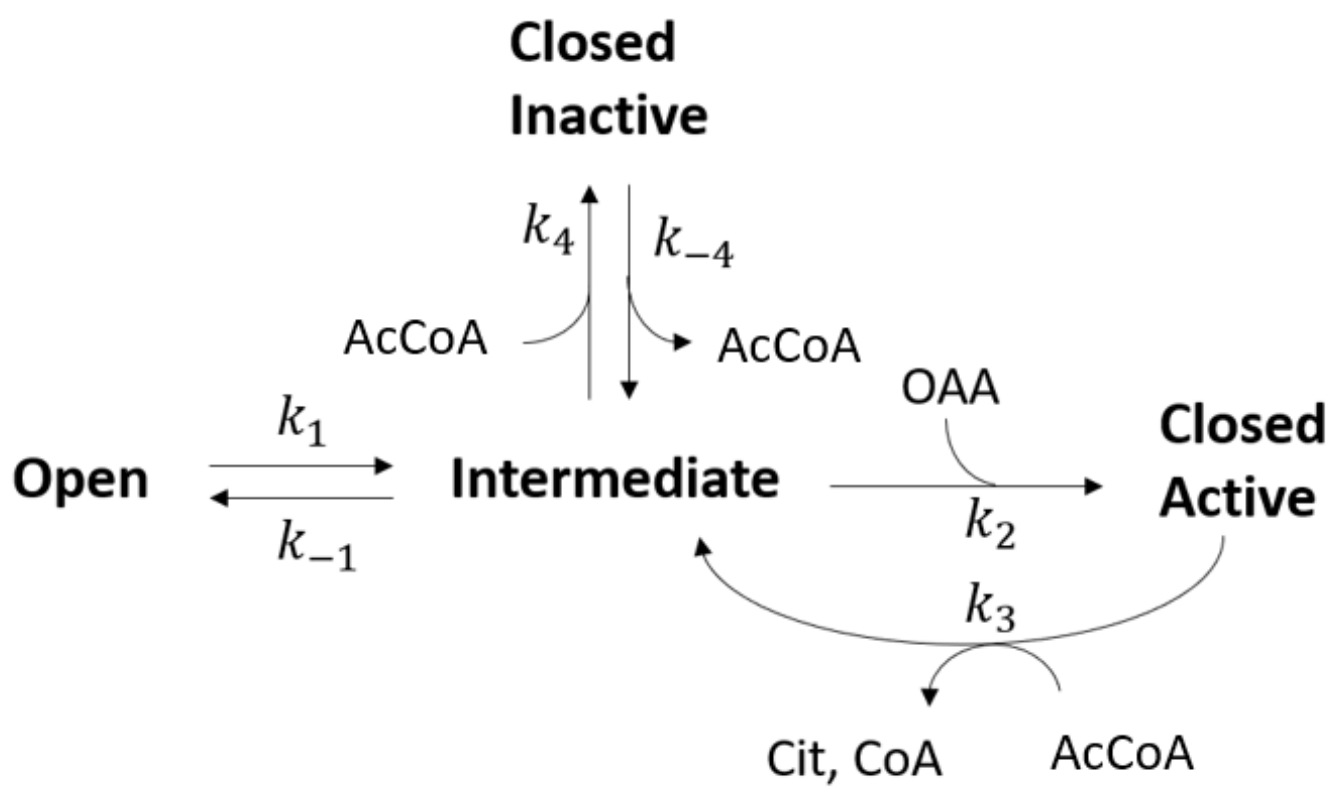

**Figure S13.** Reaction scheme for citrate synthase deduced from the transmission signal from a single trapped protein.

## SI-13.4 Reaction rates of CS from single-molecule data

The reaction scheme in Fig. S13 is similar to the one proposed by Segel[42] for substrate inhibition in an ordered bi-reactant system. However, the model of Segel describes the reaction network as a function of the various species formed as reaction intermediates, while our model describes the enzyme kinetic cycle based on conformational changes. By considering each subunit of the enzyme as independent, the model yields the following set of differential equations:

$$\frac{dN_o}{dt} = -k_1 N_o + k_{-1} N_i, \tag{S13.12}$$

$$\frac{dN_{ca}}{dt} = k_2 N_i [\text{OAA}] - k_3 N_{ca} \cdot [\text{AcCoA}], \tag{S13.13}$$

$$\frac{dN_i}{dt} = k_1 N_o - k_{-1} N_i + k_{-4} N_{ci} - k_4 N_i [\text{AcCoA}] - k_2 N_i [\text{OAA}] + k_3 N_{ca} [\text{AcCoA}], \tag{S13.14}$$

$$\frac{dN_{ci}}{dt} = -k_{-4} N_{ci} + k_4 N_i [\text{AcCoA}], \tag{S13.15}$$

$$\frac{d[\text{Cit}]}{dt} = \frac{d[\text{CoA}]}{dt} = v = k_3 N_{ca} [\text{AcCoA}], \tag{S13.16}$$

where $N_j$ represents the number of enzyme subunits in conformation $j = i, o,$ ci, ca, the square brackets indicate concentration and $k_1, k_2, k_3, k_4, k_{-1}$ and $k_{-4}$ are kinetic constants for different reaction steps as indicated in Figure S13.

In the situation where [OAA] and [AcCoA] are much higher than the enzyme concentration, we assume that the system quickly reaches a steady state with $\frac{dN_o}{dt} = \frac{dN_{ca}}{dt} = \frac{dN_i}{dt} = \frac{dN_{ci}}{dt} = 0$. These conditions well describe the plasmonic optical trapping experiment where there is a single trapped enzyme molecule and the concentration of the substrates is held constant by flow while the products are removed. With this assumption and imposing the conservation of the number of subunits $N_S = N_o + N_i + N_{ca} + N_{ci}$, we solved Equations S13.12 - S13.16 to obtain the steady state values $\overline{N_o}, \overline{N_{ca}}, \overline{N_i}$, and $\overline{N_{ci}}$

$$\overline{N_o} = \frac{k_o [\text{AcCoA}] N_S}{k_a [\text{OAA}] + (k_o + k_i)[\text{AcCoA}] + k_b [\text{AcCoA}]^2}, \tag{S13.17}$$

$$\overline{N_{ca}} = \frac{k_a[\text{OAA}]N_S}{k_a[\text{OAA}]+(k_o+k_i)[\text{AcCoA}]+k_b[\text{AcCoA}]^2}, \tag{S13.18}$$

$$\overline{N_{\text{I}}} = \frac{k_i[\text{AcCoA}]N_S}{k_a[\text{OAA}]+(k_o+k_i)[\text{AcCoA}]+k_b[\text{AcCoA}]^2}, \tag{S13.19}$$

$$\overline{N_{\text{ci}}} = \frac{k_b[\text{AcCoA}]^2 N_S}{k_a[\text{OAA}]+(k_o+k_i)[\text{AcCoA}]+k_b[\text{AcCoA}]^2}, \tag{S13.20}$$

with

$$k_a = k_1 k_2 k_{-4}\,,\; k_b = k_1 k_2 k_4\,,\; k_o = k_3 k_{-1} k_{-4}\,,\; k_i = k_1 k_3 k_{-4}. \tag{S13.21}$$

In a bulk experiment the probability $\varphi_j$ to find a protein in a certain conformation $j$ corresponds to the fraction of molecules in that conformation:

$$\varphi_j = \frac{N_j}{N_S}. \tag{S13.22}$$

However, such quantities are not meaningful in the context of single-molecule experiments. To circumvent this problem, we assume that the probability to sample a subunit of the protein in a certain conformation during a bulk experiment is equal to the fraction of time spent by each subunit in that conformation (provided that the total experiment time $t_{tot}$ is much longer than the typical residence time, $t_j$, of the protein in that conformation). Therefore,

$$\varphi_j = \frac{N_j}{N_S} = P_j = \frac{t_j}{t_{\text{tot}}}, \tag{S13.23}$$

where $P_j$ is the probability to sample the configuration $j$ from single-molecule data. Hence

$$N_j = N_S\, P_j. \tag{S13.24}$$

Substituting Equation S13.24 into Equation S13.19 we obtain the probability of intermediate state ($P_{\text{i}}$), the probability of closed state ($P_{\text{c}}$), the probability of open state ($P_{\text{o}}$),

$$P_{\text{i}} = \frac{k_i[\text{AcCoA}]}{k_a[\text{OAA}]+(k_o+k_i)[\text{AcCoA}]+k_b[\text{AcCoA}]^2}, \tag{S13.25}$$

$$P_{\text{c}} = P_{\text{co}} + P_{\text{ci}} = \frac{k_a[\text{OAA}]+k_b[\text{AcCoA}]^2}{k_a[\text{OAA}]+(k_o+k_i)[\text{AcCoA}]+k_b[\text{AcCoA}]^2}, \tag{S13.26}$$

$$P_o = 1 - P_c - P_i = \frac{k_o[\text{AcCoA}]}{k_a[\text{OAA}]+(k_o+k_i)[\text{AcCoA}]+k_b[\text{AcCoA}]^2}, \tag{S13.27}$$

Now we can rewrite Equation S13.16 by means of Equation S13.18 and obtain:

$$\frac{d[\text{Cit}]}{dt} = v = \frac{k_3 k_a[\text{OAA}][\text{AcCoA}]N_S}{k_a[\text{OAA}]+(k_o+k_i)[\text{AcCoA}]+k_b[\text{AcCoA}]^2}, \tag{S13.28}$$

Using Equation S13.25, considering that the number of enzyme molecules is $N_{\text{E}} = N_S/2$ and rearranging we can finally show that

$$v = \frac{k_2 N_{\text{E}}}{2}[\text{OAA}]P_i, \tag{S13.29}$$

Hence, the reaction rate at steady state for citrate synthase is directly proportional to the probability of sampling the intermediate conformation *i* in a single-molecule experiment.

The probabilities of sampling citrate synthase in the different conformations in the presence of 150 μM OAA and various AcCoA concentrations are globally fitted using Equations S13.25 - S13.28 and shown in Figure S14a. In agreement with previous reports of substrate inhibition of CS by AcCoA, the reaction rate, proportional to $P_i$, reaches a maximum for [AcCoA] ≈ 100 μM and then decreases.

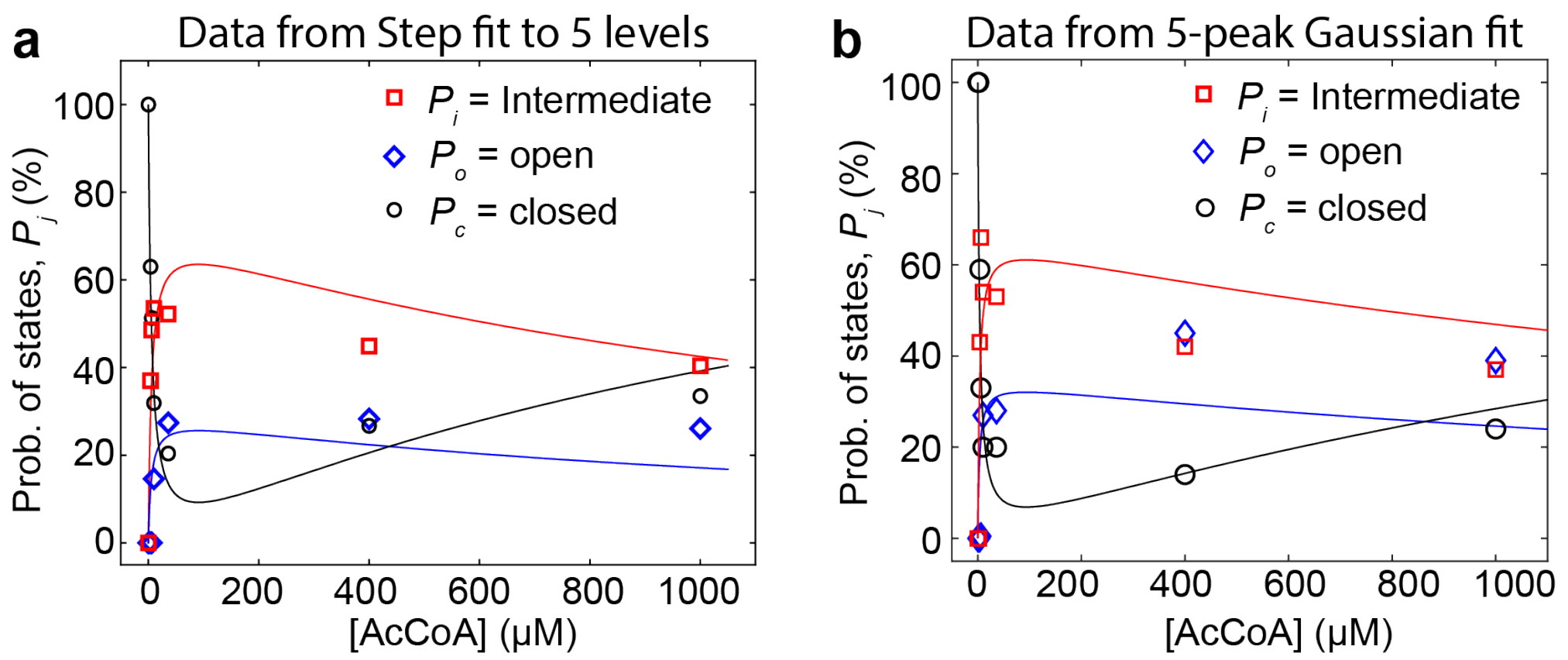

**Figure S14**. Probabilities to find a citrate synthase subunit in the *closed* ($P_c$), *open* ($P_o$) and *intermediate* ($P_i$) conformations in the presence of 150 μM OAA and at different AcCoA concentrations. **a,** Markers are the probability values calculated from the experimental data using Equations S13.1 - S13.4 (listed in Table S3). Here $P_i$ is proportional to the rate of citrate production, as shown by Equation S13.29. **b**, Markers are the values calculated from the experimental data from 5-peak Gaussian fitting to the optical transmission traces after low-pass digital filtering at 100 kHz (Table S2). The continuous curves are the global fitting to Equations S13.25 - S13.27.

To give an estimate of the apparent $K_M$, we derive an expression for the concentration of AcCoA at which the CS operates at half of its maximum speed. We define this parameter $\kappa_M$. We first calculate [AcCoA] at which the enzyme activity is maximum,

$$\frac{dv}{dt} = 0 \rightarrow [\text{AcCoA}]_{v_{\text{max}}} = \sqrt{\frac{k_a[\text{OAA}]}{k_b}} \quad \text{(S13.30)}$$

Substituting [AcCoA] with $[\text{AcCoA}]_{\text{max}}$ in Equation S13.28 yields the maximum rate $v_{\text{max}}$. By solving the equation $v = v_{\text{max}}/2$ for [AcCoA], we obtained $\kappa_M$:

$$\kappa_M = \frac{(k_O+k_I)[\text{AcCoA}]+4\sqrt{k_a k_b[\text{AcCoA}]^3}-\sqrt{[\text{AcCoA}]^2\left(\left(k_O+k_I+4\sqrt{k_a k_b[\text{AcCoA}]}\right)^2-4k_a k_b[\text{AcCoA}]\right)}}{2[\text{AcCoA}]k_b}. \quad \text{(S13.31)}$$

Equation S13.30 led to a $\kappa_M$ of 4.4 μM from the global fitting shown in Figure S14a.

Applying this kinetic model to the conformation states obtained from 5-peak Gaussian fits to 100 kHz data (see Figure 4c in the main text and Table S2), we obtained a $\kappa_M$ of 3 μM from the global fitting shown in Figure S15b and Figure 4e in the main text.

## SI-14 Bandwidth of the plasmonic optical trapping system

The recording bandwidth of the Avalanche Photodiode (APD) detector of 50 MHz potentially allows the plasmonic optical trapping setup to explore kinetic processes that occur at nanosecond time scales. To estimate the time resolution of the system, we recorded the APD signal by using an oscilloscope at a sampling rate of 500 MHz, which is ten times higher than the bandwidth of the APD. Figure S15 shows the power spectral densities (PSDs) of APD signals recorded by the oscilloscope. The presence of a trapped protein increased the PSD at frequencies less than 50 kHz. No significant difference was observed at frequencies larger than 100 kHz, indicating the absence of a high frequency noise source in the system, which makes it possible to analyze this high bandwidth signal. The decrease of the PSDs after 50 MHz (marked by the black line) result from the bandwidth limit of the APD.

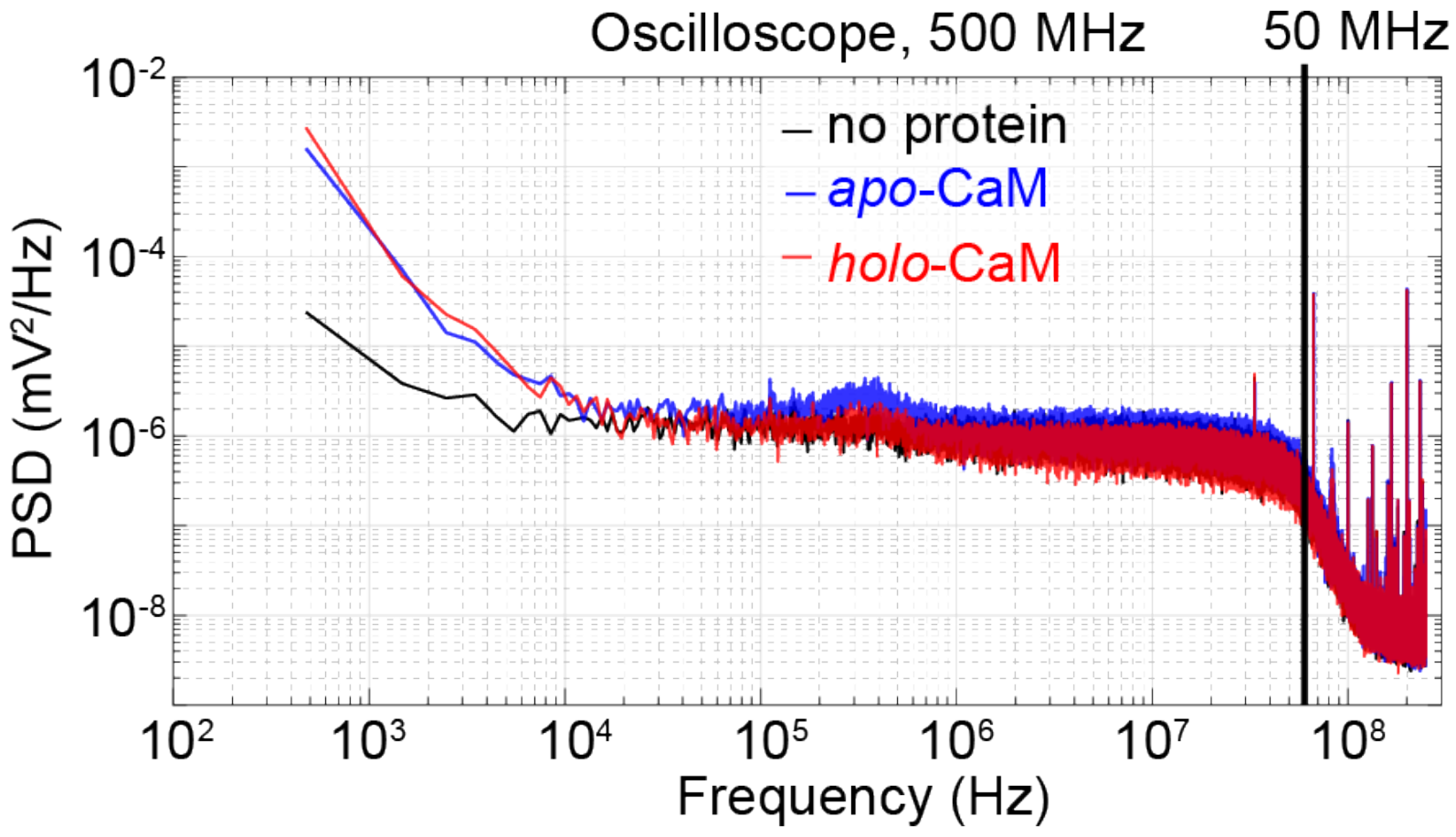

**Figure S15.** Comparison of power spectral densities of the APD signal for the same DNH in the absence of protein (black), and with protein trapped at different buffer conditions (blue and red) with a laser intensity of 23 mW. Data were recorded by an oscilloscope at a sampling rate of 500 MHz.

To test the bandwidth of the plamonic optical trapping system, we recorded the transmission signal at 15 MHz, which is the highest sampling rate of the fastest DAQ card used in this system. The transmission traces of a citrate synthase (CS) molecule trapped in a DNH (Figure S16) reveal conformational fluctuations in the raw data, with a bandwidth of approximately 7.5 MHz according to the Nyquist criterion.[43]

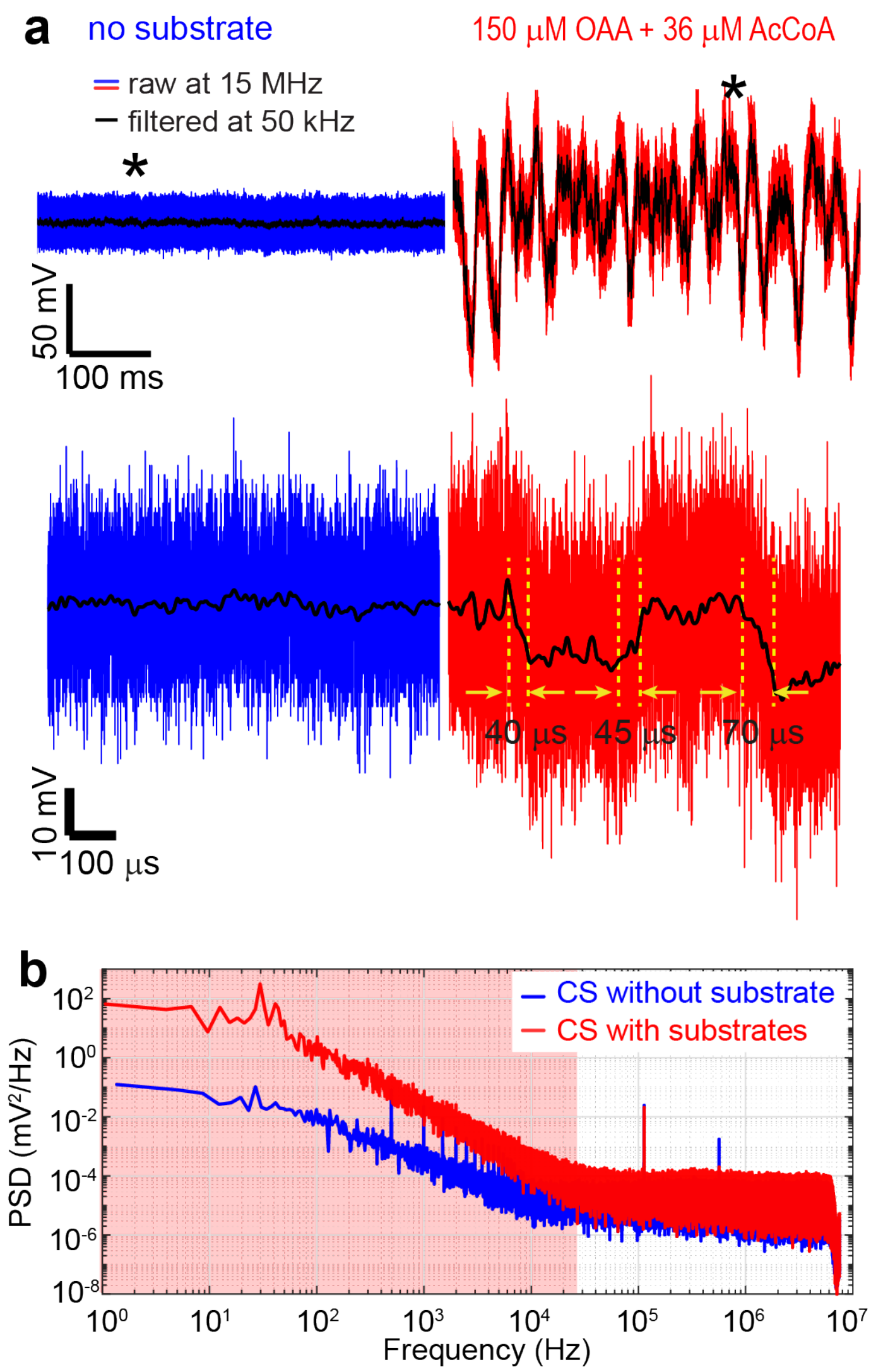

**Figure S16.** Time resolution of the plasmonic optical trapping setup. **a**, Representative transmission traces of citrate synthase (CS) trapped in a DNH structure shown both raw (colored, sampling rate of 15 MHz) and digitally filtered with a cutoff frequency of 50 kHz (black). The transmission trace for CS without substrate was recorded in a buffer of 100 mM KCl, 50 mM Tris, and 0.3 mM DTNB at pH 8.0. The transmission trace for CS with substrates was recorded in the same buffer containing 150 μM oxaloacetate and 36 μM acetyl coenzyme A. The zoomed transmission traces with expanded time scale show that CS transitioned between substrate-induced changes in transmission levels within 40 to 70 μs. These transitions were the fastest we could find although the 50 kHz low-pass filtered recording should, in theory, be able to resolve transitions as fast as approximately 20 μs. We confirmed that the fastest transition shown here still took 40 μs when the same transmission trace was low-pass filtered with a cutoff of 500 kHz instead of 50 kHz, resulting in a temporal resolution of approximately 2 μs). These results indicate that the large domain motions of CS that led to the substrate-induced changes of transmission levels occurred typically within 40 μs or slower and these results also show that these substrate-induced changes in CS conformation could be monitored with a temporal resolution at least as fast as 40 μs and possibly faster. **b**, Comparison of power spectral densities (PSDs) of the 15 MHz traces shown in panel a. Adding the two substrates induced an increase of the PSD at the frequency range below 50 kHz, as indicated by the red area. No clear difference at the frequencies higher than 100 kHz was observed.